\documentclass[12pt]{article}
\usepackage{graphicx}
\usepackage{amssymb}
\usepackage{amsmath}
\usepackage{xcolor}
\usepackage{soul}
\usepackage{cite}
\usepackage{caption}

\usepackage{mathtools}
\DeclarePairedDelimiter\ceil{\lceil}{\rceil}

\usepackage{array}
\usepackage{subcaption}
\usepackage{booktabs}

\makeatletter

\@addtoreset{equation}{section}
\makeatletter

\DeclareMathOperator{\sgn}{sgn}

\setstcolor{red}

\begin{document}

\begin{titlepage}
\vfill
\begin{flushright}
\end{flushright}

\begin{center}
\baselineskip=16pt
{\Large\bf 
Overspinning naked singularities \\ in AdS$_3$ spacetime\\
}
\vskip 0.5cm
\vskip 10.mm
{\bf Mat\'{\i}as Brice\~no${}^{a}$, Cristi{\'a}n Mart\'{\i}nez${}^{b}$ and Jorge Zanelli${}^{b}$} \\
\vskip 1cm
{${}^a$ Instituto de F\'{\i}sica, Pontificia Universidad Cat{\'o}lica de Chile, Casilla 306, Santiago, Chile\\
${}^b$ Centro de Estudios Cient\'{\i}ficos (CECs), Av. Arturo Prat 514, Valdivia, Chile \\
\vskip 0.2cm
\texttt{\footnotesize{matias.briceno.97@gmail.com, martinez@cecs.cl, z@cecs.cl}}
} 
\vspace{6pt}
\end{center}
\begin{center}
{\bf Abstract}
\end{center}
The  BTZ black hole belongs to a family of locally three-dimensional anti-de Sitter (AdS$_3$) spacetimes labeled by their mass $M$ and angular momentum $J$. The case $M \ell \geq |J|$, where $\ell$ is the anti-de Sitter radius, provides the black hole. Extending the metric to other values of of $M$ and $J$ leads to geometries with the same  asymptotic behavior and global symmetries, but containing a naked singularity at the origin. The case $M \ell \leq -|J|$ corresponds to spinning conical singularities that are reasonably well understood. Here we examine the remaining case, that is $-|J|<M\ell<|J|$. These naked singularities are mathematically acceptable solutions describing classical spacetimes. They are obtained by identifications of the covering pseudosphere in  $\mathbb{R}^{2,2}$ and are free of closed timelike curves. Here we study the causal structure and geodesics around these \textit{overspinning} geometries. We present a review of the geodesics for the entire BTZ family. The geodesic equations are completely integrated, and the solutions are expressed in terms of elementary functions. Special attention is given to the determination of circular geodesics, where new results are found. According to the radial bounds, eight types of noncircular geodesics appear in the BTZ spacetimes.  For the case of overspinning naked singularity, null and spacelike geodesics can reach infinity passing by a point nearest to the singularity, others extend from the central singularity to infinity, and others still have a radial upper bound and terminate at the singularity. As expected for an anti-de Sitter spacetime, timelike geodesics cannot reach infinity; they either loop around the singularity or fall into it. The spatial projections of the geodesics (orbits) exhibit self-intersections, whose number is determined for null and spacelike geodesics, and it is found a special class of timelike geodesics whose spatial projections are closed.
\end{titlepage}

\section{Introduction} 
 
All vacuum solutions of the three-dimensional Einstein equations with negative cosmological constant $-\ell^{-2}$ are spacetimes of constant negative curvature, locally isometric to AdS$_3$. This extremely simple classification of all possible local geometries allows for a variety of spacetimes with radically different global structures. Inequivalent physical configurations such as black holes and naked conical singularities are locally indistinguishable but globally very different from AdS$_3$. Even specifying global symmetries allows for completely unrelated geometries. For instance, as shown in \cite{AyonBeato:2004if}, locally constant curvature cyclic symmetric spacetimes --namely, those with an $SO(2)$ isometry characterized by a globally defined Killing vector-- are the BTZ \cite{BTZ1,BTZ2}, the self-dual Coussaert-Henneaux spacetimes \cite{C-H} and the toroidal time-dependent geometries \cite{AyonBeato:2004if}, with isometry groups $SO(2)\times\mathbb{R}$, $SO(2)\times SO(2,1)$ and $SO(2)\times SO(2)$, respectively.

 The BTZ family of spacetimes is described by the stationary line element\footnote{We set the three-dimensional Newton constant as $G=1/8$.}
\begin{equation} \label{BTZmetric}
g_{\mu \nu} dx^{\mu}dx^{\nu}= -\left(\frac{r^2}{\ell^2}-M\right)dt^2 -Jdt d\theta + \left(\frac{r^2}{\ell^2}-M+\frac{J^2}{4r^2}\right)^{-1}dr^2+r^2 d\theta^2,
\end{equation}
where the coordinate ranges are: $-\infty<t<\infty$, $0<r<\infty$, and $0\leq \theta\leq 2\pi$. Here the mass $M$ and angular momentum $J$ are the conserved charges associated to the Killing vectors $\partial/\partial t$ and $\partial/\partial \theta$, respectively. Different values of $M$ and $J$ lead to distinct spacetimes. A quick route to recognize the spacetimes described by \eqref{BTZmetric} is to study the roots of $g^{rr}=0$, namely
\begin{equation}
\frac{r^4}{\ell^2} - M r^2 + \frac{J^2}{4} =0 .
\end{equation}
Since this equation is a function of $r^2$, the four roots take the form $\{\lambda_{\pm}, \, -\lambda_{\pm}\}$, with
\begin{equation} \label{roots}
\lambda_{\pm}=\frac{\ell}{2}\left[\sqrt{M+J/\ell}\pm \sqrt{M-J/\ell}\right].
\end{equation}
Of these, at most two are real and positive which occurs for $M\ell \ge |J|$, corresponding to the horizons of a rotating BTZ black hole. This 2+1 dimensional black hole shares most of the attributes of the Kerr solution in 3+1 dimensions, and like its more realistic counterpart, contains an inner horizon and also an ergoregion. 

\begin{figure}[h!]
    \centering
    \includegraphics[scale=0.7]{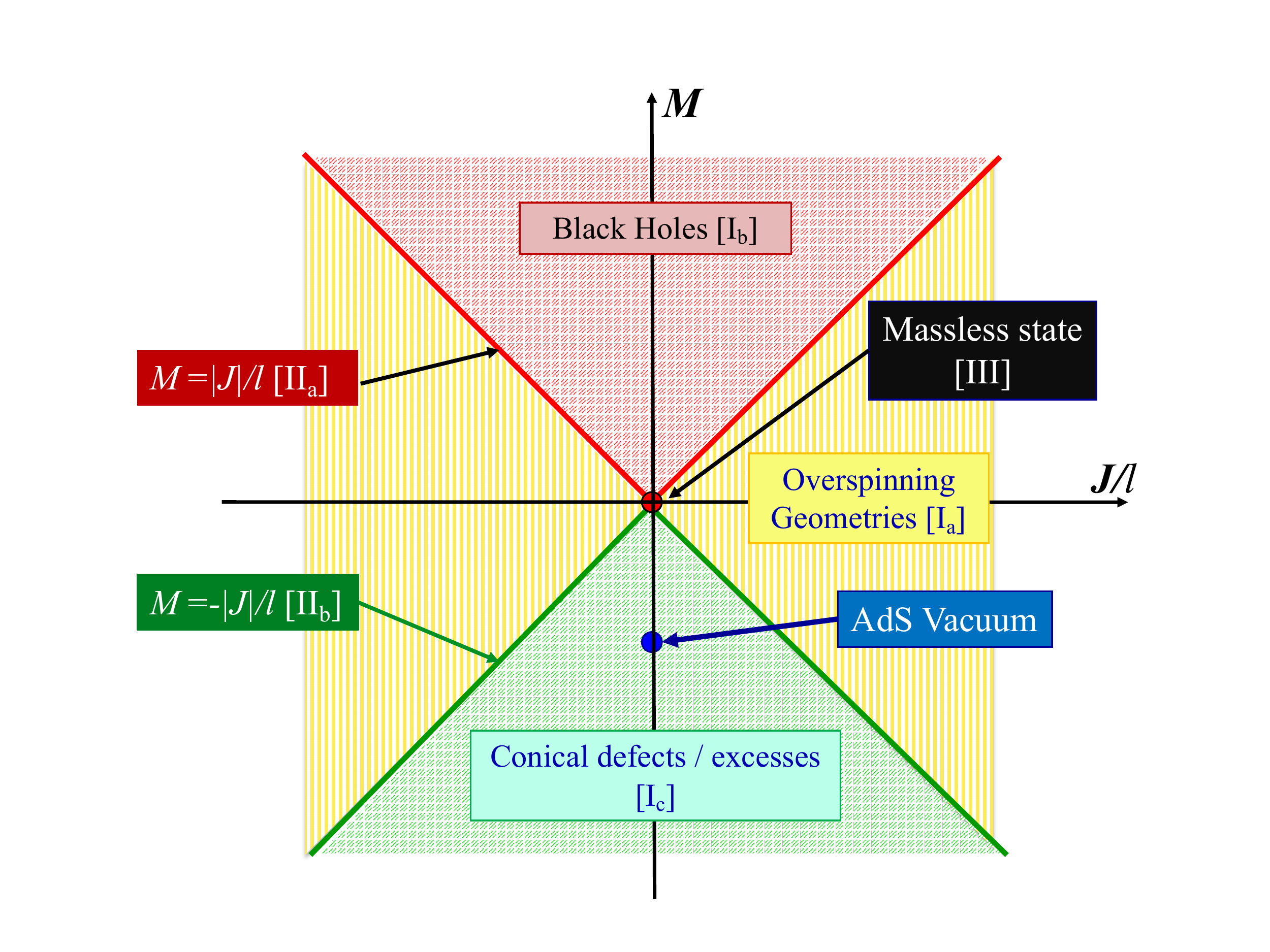}
    \caption{BTZ spacetimes for all ranges of $J$ and $M$. The labels in brackets correspond to the classification of \cite{BTZ2} reproduced in Table \ref{tablevectors}.}
    \label{MonoX}
\end{figure}

Other regions in the $M-J$ plane lead to complex roots which correspond to different geometries, as depicted in Fig. \ref{MonoX} and summarized in Table  \ref{tablevectors}. In this article, our attention is focused on the case $J^2>M^2\ell^2$, for which the values of $\lambda$ are complex, leading to rotating spacetimes with a naked singularity at $r=0$. Surprisingly, although this solution was pointed out almost 30 years ago in \cite{BTZ2}, it has not been discussed much so far. Since in this case the angular momentum is greater than the mass, we will refer to it as the \textit{overspinning} spacetime. 

As is well known, BTZ geometries can be obtained through identifications in the universal covering space CAdS$_3$. The construction of the black hole was presented in \cite{BTZ2}, including the classification of all one-parameter subgroups of $SO(2,2)$ with the associated Killing vectors and Casimir invariants. A similar construction for the case of the conical spacetime can be found in \cite{MZ}. The Killing vectors for the identifications corresponding to the different geometries are shown in Table \ref{tablevectors}.

The main purpose of this work is to examine the overspinning spacetime, which, like the rest of BTZ geometries, is free from closed timelike curves, but contains a naked singularity of different nature from that of the conical defects or excesses \cite{BMZ2}. The analysis is carried out in a setup that includes the other members of the BTZ family, which allows for a direct comparison among the different spacetimes including a review of previous results. As a first step, the identification in CAdS$_3$ that leads to the overspinning geometry is explicitly displayed and the nature of the singularity is briefly discussed in Sec. \ref{construction}. In Sec. \ref{sec3}, a review of general features of the geodesics --timelike, spacelike and null-- for the metric \eqref{BTZmetric} is presented. The analysis considers generic and circular geodesics which complements the results presented in \cite{Cruz1994} for $M\ell>|J|$ (black holes), and in \cite{Martinez:2019nor} for $M\ell<-|J|$ (conical singularities). This section also includes general properties of the geodesics for the remaining case $|M|\ell<|J|$ (overspinning singularities). The explicit expressions and specific behavior of the geodesics for the overspinning spacetime are discussed in Sec. \ref{GONS}.
Finally, Sec. \ref{sec5} contains a summary of results and outlines the road ahead.

\section{Nature of the overspinning geometry} \label{construction} 
We now briefly revisit the construction of BTZ geometries described by \eqref{BTZmetric} through identifications on the universal covering space CAdS$_3$, as shown in \cite{BTZ2}. We will focus on the overspinning geometry, singled out by the condition $J^2>M^2\ell^2$.

\subsection{Construction as an identification on CAdS$_3$} \label{identificaciones}
Consider CAdS$_3$, the three-dimensional constant negative curvature manifold defined as the set of points with coordinates  $X^A=\left(X^{0},X^{1},X^{2},X^3\right)$ in $\mathbb{R}^{2,2}$, given by\footnote{Here $X_A=\eta_{AB}X^B$, with $\eta_{AB}=\text{diag}(-,-,+,+)$. In terms of the coordinates chosen in \cite{BTZ2}, $X^{A}=(v,u,x,y)$.} 
\begin{equation} \label{pseudosphere}
\eta_{AB} X^A X^B = -(X^0)^2-(X^1)^2+(X^2)^2+ (X^3)^2=-\ell^2. 
\end{equation}
This pseudo-sphere is a maximally symmetric manifold with six globally defined Killing vectors, corresponding to the six generators of $so(2,2)$,
\begin{equation} \label{generators}
J_{A B}:=X_B \partial_A-X_A \partial_B .
\end{equation}
Identifying points on this surface related by a Killing vector turns this hypersurface in $\mathbb{R}^{2,2}$ into a different manifold with the same local geometry. The pseudo-sphere \eqref{pseudosphere} can be covered with a system of coordinates, which in the induced metric \eqref{BTZmetric} is given by $(t, r, \theta)$. The identifying Killing vector $\Theta$ in the original hypersurface becomes $\partial_{\theta}$, which is also a Killing vector in the resulting manifold. 

The explicit form of the embedding for black holes and spinning conical spacetimes can be found in  \cite{BTZ2} and \cite{MZ}, respectively.\footnote{This construction has been recently reviewed in \cite{CFMZ} for all cases satisfying $M^2\ell^2\ge J^2$.} In what follows, we will determine the embedding for the overspinning spacetime. The Killing vector $\Theta$ is a linear combination of the $so(2,2)$ generators of the form
\begin{equation} \label{genKilling}
\Theta= \frac{1}{2} \omega^{A B} J_{A B} =\omega^{A B} X_{B} \partial_{A}\,,
\end{equation} 
where the antisymmetric matrix $\omega^{A B}$ characterizes the identification.\footnote{Note that the matrix $H := e^{2\pi \Theta}$, acting as $H^{A}\,_{B} \, X^B(t, r, \theta) = X^A(t, r,\theta +2\pi)$ in the embedding space, amounts to a rotation by $2\pi$ in the resulting manifold.} A complete classification of $\omega^{A B}$ up to conjugation given in \cite{BTZ2} provides the canonical forms for the antisymmetric matrices, which in turn yields different spacetimes. This classification is summarized in Table \ref{tablevectors} following the notation of \cite{BTZ2}.

\begin{table}[ht!] 
\centering
\caption{Roots of $g^{rr}=0$ and identification Killing vectors $\Theta$ in terms of $so(2,2)$ generators for different BTZ geometries. The types are shown in bold following the notation of \cite{BTZ2}.}
\scalebox{0.9}{
\begin{tabular}{c|c|c}
\hline
 \bf{Roots of $g^{rr}=0$} & \bf{Killing vector $\boldsymbol{\Theta}$} & \bf{Geometry} \\[1.5mm] \hline \hline 
$\lambda_+\,, \lambda_{-} \in \mathbb{R}$ & $\lambda_+ J_{0 3}+\lambda_- J_{12}$ & Generic BH,  $M\ell > |J|$ \bf{(I$_{b}$)} \\[1.5mm]
\hline 
$\lambda_{+}=|\lambda_{-}| \in \mathbb{R}^+$ & $\lambda_+(J_{03}\pm J_{12})+\frac{1}{2}(J_{02}+J_{01}+J_{32}+J_{31})$& Extremal BH, $M\ell = |J|$ \bf{(II$_{a}$)} \\[1.5mm]
\hline 
 $\lambda_+\,, \lambda_{-} \in i\mathbb{R}$ & $-i\lambda_- J_{23}-i \lambda_+ J_{01}$ & Generic CS, $M\ell < -|J|$ \bf{(I$_{c}$)}\\[1.5mm] \hline 
$\lambda_+=\pm \lambda_- \in i\mathbb{R}$ & $i\lambda_+(J_{01}+J_{23})-\frac{1}{2} (J_{03}+J_{01}+J_{32}+J_{12})$ & Extremal CS, $M\ell = -|J|$ \bf{(II$_b$)} \\[1.5mm] \hline 
$\lambda_{\pm}=0$ & $J_{32}-J_{31}$  & Massless geometry, $M = 0 = J$ \bf{(III)}\\[1.5mm] \hline
$\lambda_{\pm}=a \pm ib$ & $b(J_{01}+J_{23})-a(J_{03}+J_{12})$ & Overspinning spacetime, \\  $a, b \in \mathbb{R}$, see (\ref{ab})  &  & $-|J|< M\ell < |J|$ \; \bf{(I$_{a}$)} \\ [1.5mm] \hline
\end{tabular} }\label{tablevectors}
\end{table}

As shown in Table \ref{tablevectors}, all geometries of the family (\ref{BTZmetric}) are obtained by Killing vector identifications in the covering of AdS$_3$. The non-extremal black hole ($M>|J|/\ell$) results from identifying along a Killing vector formed by a linear combination of two boosts, while the non-extremal conical singularity is obtained identifying by two spatial rotations. The identification for the overspinning spacetime is a linear combination of a rotation and a boost. Note that the Killing vectors for the extremal and massless cases are not limiting cases of the generic forms.

The canonical form of the Killing vector associated to overspinning spacetime is extracted from Table \ref{tablevectors} and it reads (in the basis $\partial_A$) as 
\begin{align}\label{ONSKilling}
    \Theta=(-a X^3-b X^1, b X^0-a X^2, b X^3-a X^1, -a X^0-b X^2), 
\end{align}
where 
\begin{equation}\label{ab}
 a=\frac{\sqrt{|J|/\ell+ M}}{2},\quad b= \frac{\sqrt{|J|/\ell- M}}{2}.
\end{equation} 
The matrix $\omega^{A}\,_{B}$ that characterizes the above Killing vector is then given by 
\begin{align}\label{omegaK}
     \omega=\begin{pmatrix} 0 && -b && 0 && -a \\ b && 0 && -a && 0\\ 0 && -a && 0 && b\\ -a && 0 && -b && 0 \end{pmatrix}.
\end{align}
Note that the Casimir invariants  $I_1=\omega_{AB} \omega^{A B}$ and $I_2=\frac{1}{2}\epsilon_{ABCD} \omega^{AB}  \omega^{CD}$ become $I_1=4(b^2-a^2)= -2M$ and $I_2=4(b^2+a^2)= 2|J|/\ell$, respectively. They satisfy $I_1+I_2>0$ and $I_1-I_2<0$, consistent with the fact that $\omega^{A B}$ is an antisymmetric tensor of type I$_{a}$ \cite{BTZ2}.

Since $\Theta$ is just $ \partial_{\theta} =\displaystyle \frac{\partial X^A}{\partial \theta}\partial_{A}$, from Eq. \eqref{genKilling} we get
\begin{align}\label{partialeq}
    \frac{\partial X^{A}}{\partial \theta} = \omega^{A}\,_{B} X^{B}.
\end{align}
Solving this equation we obtain the $\theta$-dependence of $X^A$. The explicit dependence on $r$ and $t$ is determined from \eqref{pseudosphere} and by matching \eqref{BTZmetric} with the induced metric
\begin{align}\label{4D_metric}
    ds^2 =\eta_{AB} dX^A dX^B = - (dX^0)^2 - (dX^1)^2 + (dX^2)^2 + (dX^3)^2,
\end{align}
For $J>0$, the explicit form of the embedding is thus found to be 
\begin{align} \label{eJ>0}
 X^0&=\frac{\ell}{2}  \sqrt{A(r)-1} \sinh [a (t/\ell -\theta)]\{\sin[b (t/\ell +\theta)] + \cos[b (t/\ell +\theta)] \} \nonumber \\&+\frac{\ell}{2}  \sqrt{A(r)+1} \cosh [a (t/\ell -\theta)]\{\cos[b(t/\ell + \theta)]-\sin[b(t/\ell + \theta)]\}, \nonumber \\
 X^1&=\frac{\ell}{2}  \sqrt{A(r)+1} \cosh[a (t/\ell - \theta)]\{\sin[b(t/\ell + \theta)] + \cos[b(t/\ell + \theta)]\}\nonumber \\&-\frac{\ell}{2} \sqrt{A(r)-1} \sinh[a(t/\ell - \theta)]\{\cos[b(t/\ell + \theta)]-\sin \left[b \left(t/\ell + \theta \right)\right]\}, \nonumber \\
 X^2 &=\frac{\ell}{2} \sqrt{A(r)+1} \sinh[a (t/\ell - \theta)] \{\sin[b(t/\ell + \theta)] + \cos [b(t/\ell + \theta )]\}\nonumber \\&-\frac{\ell}{2}  \sqrt{A(r)-1} \cosh[a(t/\ell - \theta )]\{\cos[b(t/\ell + \theta)] - \sin[b (t/\ell + \theta)]\}, \nonumber\\
X^3&=\frac{\ell}{2} \sqrt{A(r)+1} \sinh[a (t/\ell - \theta)]\{\cos[b(t/\ell + \theta)] - \sin[b(t/\ell + \theta)]\}\nonumber \\&+\frac{\ell}{2}\sqrt{A(r)-1} \cosh[a (t/\ell - \theta)]\{\sin[b(t/\ell + \theta)] + \cos[b(t/\ell + \theta)]\},
\end{align}
with
\begin{equation}\label{A(r)}
A(r)=\frac{2 \sqrt{\frac{J^2}{4}+\frac{r^4}{\ell^2}-M r^2}}{\sqrt{J^2-\ell^2M^2} }.
\end{equation}
The embedding for $J<0$ is obtained replacing $t$ by $-t$ in \eqref{eJ>0}.
Unlike the case of the black hole, only one patch is required to cover the whole overspinning spacetime given by \eqref{eJ>0}. Indeed, it can be seen that $A(r)\pm 1 > 0 $ since
\begin{align}
    A^2(r)=\frac{4(\frac{r^2}{\ell}-\frac{M \ell}{2})^2 + J^2 -M^2 \ell^2}{J^2-\ell^2 M^2} =\frac{4(\frac{r^2}{\ell}-\frac{M \ell}{2})^2}{J^2-\ell^2 M^2} +1.
\end{align}

The original pseudo-sphere \eqref{pseudosphere} is invariant under $SO(2,2)$ and therefore possesses 6 globally defined Killing vectors. After the identification, only two Killing vector fields remain, namely, $\partial_t$ and $\partial_\theta$. Note that  for $J \neq 0$ the identification has no fixed points. Only in the static geometries ($J=0$) the Killing vector has a single fixed point at $r=0$ \cite{BTZ2,Steif}.

The surface $r=0$ ($\forall \, t$) is not a coordinate singularity. It is the central singularity of black holes, conical singularities and overspinning geometries,  which is irremovable. The nature of this singularity can be explored by parallel transport of a vector in a closed loop around $r=0$. For each of those geometries, in the limit of zero radius for the loop, the parallel transport produces a vector rotated by a finite element of the Lorentz group \cite{BMZ2}. Hence, the surface $r=0$ could be described as the support of a Dirac delta singularity in the Lorentz curvature, analogous to the finite rotation produced at the tip of a cone. Therefore, at $r=0$ the tangent space is not properly defined and those points cannot be genuinely considered as part of the manifold. Nevertheless, the tangent space is well defined on every open set of the manifold that does not include $r=0$, which is a boundary beyond which the spacetime manifold is not defined. In particular, geodesics that reach $r=0$ cannot be extended and consequently geometries described by \eqref{BTZmetric} are geodesically incomplete \cite{Hawking-Ellis}.

It must be stressed that although the geometries with $M \ell< |J|$  contain a naked singularity, one can still investigate physically meaningful problems like the motion of test particles or the propagation of signals. This is because these naked singularities belong to a mild class of \textit{quasi-regular} singularities that behave as boundaries \cite{Ellis-Schmidt}. Other naked singularities, such as the one in the Schwarzschild geometry for $M<0$ are physically intractable because the curvature scalars diverge as $r\to 0$ and the initial value problem could not be meaningfully formulated in the presence of those strong naked singularities.

\subsection{Causal structure} \label{causal} 

\begin{figure}[hb!]
    \centering
    \includegraphics[scale=0.2]{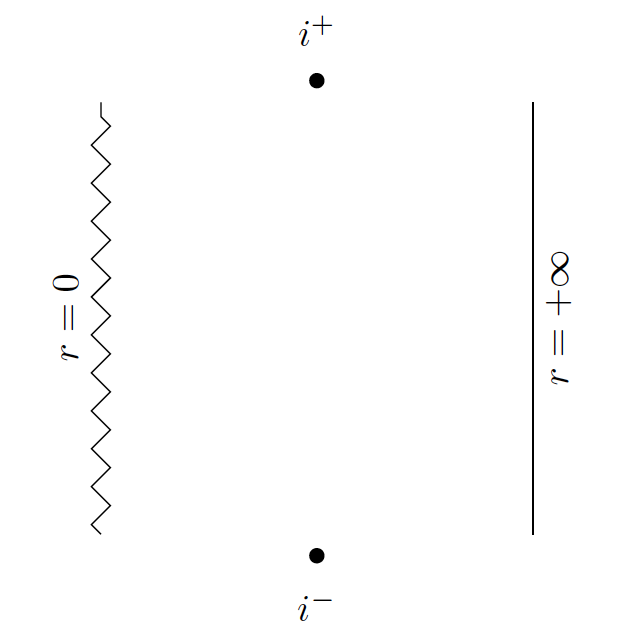}
    \caption{Penrose diagram for the overspinning geometry (see Eq.  \eqref{metric_R}).}
    \label{ads_penro}
\end{figure}

The causal structure of the different geometries described by \eqref{BTZmetric}, whose $2+1$ form is 
\begin{align}\label{Metric_ADM}
    ds^2=-(N^{\perp})^2 dt^2 + (N^{\perp})^{-2}dr^2+r^2(d\theta+ N^{\theta} dt)^2,
\end{align}
with $(N^{\perp})^2 = r^2/\ell^2 -M+ J^2/(4r^2)$ and $N^{\theta}=-J/(2r^2)$, can be studied through their Penrose diagrams. In order to take into account the effect of rotation of the spacetime properly, the Penrose diagram is drawn from the perspective of a co-rotating falling observer as in \cite{BTZ2}. A co-rotating observer has the angular velocity
\begin{align}
    \frac{d\theta}{dt}=\frac{J}{2 r^2},
\end{align}
leading to a two-dimensional Penrose diagram for the metric
\begin{align}\label{Metric_ADM2}
    ds^2_{(2)}=-(N^{\perp})^2 dt^2 + (N^{\perp})^{-2}dr^2,
\end{align}
which can also be written as
\begin{align}\label{metric_R}
    ds^2_{(2)}=\left(\frac{r^2}{\ell^2}-M+\frac{J^2}{4r^2}\right) (-dt^2+dR^2).
\end{align}
The explicit form of the new radial coordinate $R$ in terms of $r$ depends of the type of BTZ geometry. In the case of the overspinning  geometry, $R$ is given by
\begin{align}\label{compact_r}
    R =&\frac{-\ell}{4ab} \left( a\mbox{ arccot}\left( \frac{r^2/\ell^2 -a^2-b^2}{2 b r/\ell} \right) + b \log\left[\frac{\sqrt{(a^2+b^2)^2-2(a^2-b^2)r^2/\ell^2+r^4/\ell^4}}{b^2+(r/\ell-a)^2}\right] \right),
\end{align}
where $a$ and $b$ are defined in \eqref{ab}, with $0 < R < \pi\ell/(4b)$. Note that $R$ is a continuous, monotonic increasing function of $r$ and it is therefore invertible. Thus, we can construct the Penrose diagram for the overspinning geometry, shown in Figure \ref{ads_penro}. Note that  $i^+$ and $i^-$ are at infinite distance. Another feature is that $r=\infty$ and $r=0$ are timelike and there are no globally defined Cauchy surfaces. In this sense the resulting Penrose diagram shows the same causal structure as the entire AdS$_3$ spacetime, except for the singularity at $r=0$.

\subsection{Absence of closed timelike curves} \label{noCTC} 
We conclude this section by showing that the geometries described by (\ref{BTZmetric}) have no closed causal curves for any values of $M$ and $J$ for the range of coordinates indicated in \eqref{BTZmetric}. This was proven in \cite{BTZ2} for the rotating and extremal black holes where $M \geq |J|/\ell\geq 0$. We extend that argument to the remaining cases, the conical ($-M \geq |J|/\ell$) and overspinning  ($M^2 <J^2/\ell^2$) spacetimes, which have a naked singularity at $r=0$.

In order to prove the absence of closed timelike curves it is sufficient to show that there are no future-directed, timelike or null curves in the region $\Theta \cdot \Theta >0$ of the anti-de Sitter covering space that join a point and its image generated by $\exp(2 \pi \Theta)$. Consider the anti-de Sitter metric in the form
\begin{equation}
ds^2 = -(N^{\perp}(r))^2 dt^2 + (N^{\perp}(r))^{-2} dr^2 + r^2[d\theta + N^{\theta} dt]^2,
\end{equation}
where $-\infty < \theta < +\infty$. Consider a causal curve $\{t(\lambda), r(\lambda), \theta (\lambda)\}$, where the tangent vector $(dt/d\lambda, dr/d\lambda, d\theta/d\lambda)$ does not vanish for any value of $\lambda$. The curve is causal if it is timelike or null at every point,
\begin{equation}
-(N^{\perp})^2 \left(\frac{dt}{d\lambda}\right)^2 
+ (N^{\perp})^{-2} \left(\frac{dr}{d\lambda}\right)^2 
+ r^2\left(\frac{d\theta}{d\lambda} + N^\theta \frac{dt}{d\lambda} \right)^2 \leq 0.
\end{equation}

A causal curve connecting the point $(t_0, r_0, \theta_0)$ to $(t_0, r_0, \theta_0 + 2k \pi)$, must be such that for some value of $\lambda$, $dt/d\lambda =0$ because $t$ comes back to its initial value. But then, if $(N^{\perp})^2 > 0$, which holds for $-M \geq |J|/\ell$ and $M^2 <J^2/\ell^2$, it would follow that at that point $dr/d\lambda = d\theta/d\lambda = 0$, which contradicts our initial assumption.

It should be observed that if one were to admit the region $\Theta \cdot \Theta=r^2<0$ in the solution,  causality could be violated. The boundary separating the regions with and without closed causal curves is then the timelike surface $r = 0$.

In \cite{Deser:1991ye} it was argued that there are no closed timelike curves (CTCs) if the spacetime has physically acceptable global structure, which it does ``for physically acceptable sources". By physically acceptable those authors meant spacetimes that arise from regular initial conditions with normal matter. It is not clear whether the overspinning geometry can arise from a regular initial state, but the absence of CTCs means that it shares this important feature with the class of physically acceptable geometries.

\section{Geodesics in BTZ geometries} \label{sec3} 
Two conserved quantities along the geodesic lines can be obtained from the Killing vectors  $\xi=\partial_t$ and $\Theta=\partial_\theta$ of the BTZ spacetimes. They are  $E=-\xi_{\mu}\dot{x}^{\mu}$ and $L=\Theta_{\mu}\dot{x}^{\mu}$,  where $\dot{x}^{\mu}=dx^{\mu}/d\lambda$ is the velocity along the geodesic with affine parameter $\lambda$. These conserved quantities allow to write first integrals for $t$ and $\theta$,
\begin{align}
\label{t_dot}
\dot{t}&= \frac{E r^2 - JL/2}{r^2\left(\frac{r^2}{\ell^2}-M+\frac{J^2}{4r^2}\right)},\\
\label{theta_dot} 
\dot{\theta} &= \frac{(r^2/\ell^2 - M)L +J E/2}{r^2\left(\frac{r^2}{\ell^2}-M+\frac{J^2}{4r^2}\right)}.
\end{align}
Choosing the normalization $\dot{x}^{\mu} \dot{x}_{\mu}=-\varepsilon$, where $\varepsilon=1,-1,0$ for timelike, spacelike and null geodesics respectively, an additional first integral is obtained
\begin{equation}
\label{r_dot}
 r^2\dot{r}^2 = -\varepsilon r^2\left(\frac{r^2}{\ell^2}-M+\frac{J^2}{4r^2}\right)+\left(E^2-\frac{L^2}{\ell^2}\right)r^2 +L^2 M - JEL.
\end{equation}
The first-order system \eqref{t_dot}-\eqref{r_dot} determines the geodesic structure of the BTZ spacetimes. 
Although we already have obtained the first integrals, it is useful to know $\ddot{r}$. The radial component of the geodesic equation $\ddot{r}+\Gamma^{r}_{\alpha \beta}\dot{x}^{\alpha}\dot{x}^{\beta}=0$ reads
\begin{equation}
\ddot{r}=r  \left(\frac{r^2}{\ell^2}-M+\frac{J^2}{4 r^2}\right)\left(\dot{\theta}^2-\frac{\dot{t}^2}{\ell^2}\right)+\left(\frac{r^2}{\ell^2}-M+\frac{J^2}{4 r^2}\right)^{-1}\left(\frac{r^2}{\ell^2}-\frac{J^2}{4 r^2}\right)\frac{ \dot{r}^2}{r }.
\end{equation} 
Replacing \eqref{t_dot}-\eqref{r_dot} in the above equation, one obtains the second-order radial equation
\begin{equation}
\label{r_dotdot}
r \ddot{r}= -\varepsilon  \left(\frac{r^2}{\ell^2}-\frac{J^2}{4 r^2}\right)-\frac{M L^2-J E L}{r^2}.
\end{equation}

\subsection{Generic geodesics}  
In order to study the generic motion it is convenient to define a new radial variable
\begin{align} \label{defu}
u= \frac{r^2}{\ell^2} >0,
\end{align}
and the dimensionless quantities
\begin{align} \label{dimensionless}
  \tilde{t}=t/\ell, \quad    \tilde{L}=L/\ell, \quad  \tilde{J}=J/\ell,  \quad \tilde{\lambda}=\lambda/\ell.
\end{align}
Omitting the tildes from now on, the geodesic equations (\ref{t_dot}), (\ref{theta_dot}) and (\ref{r_dot}) become
\begin{align} \label{dott}
\dot{t}&= \frac{E u - JL/2}{R(u)},\\
\label{dottheta} 
\dot{\theta} &= \frac{(u - M)L +J E/2}{R(u)}, \\
\label{dotu}
\frac{\dot{u}^2}{4} &= -\varepsilon R(u)+\left(E^2-L^2\right)u+M L^2 - JEL \equiv h(u),
\end{align}
respectively, where
\begin{align}
R(u)= u^2 -M u +\frac{J^2}{4}=u g^{rr}(u), \quad \text{with} \quad u>0. 
\end{align}
Using \eqref{defu} and \eqref{dotu}, the second-order radial equation \eqref{r_dotdot} becomes
\begin{equation}\label{u_dotdot}
 \ddot{u}+4\varepsilon \, u= 2\left(\varepsilon M+E^2-L^2\right),
\end{equation}
which is a necessary condition for \eqref{dotu} valid for $u\neq 0$. The solution for this equation is a trigonometric ($\varepsilon>0$), hyperbolic ($\varepsilon<0$), parabolic ($\varepsilon=0$) or linear ($\varepsilon=0$, $E^2=L^2$) function of $\lambda$. Specifically, the radial motion is given by the positive values of 
\begin{equation} \label{ulambda}
u(\lambda)=\left\{
\begin{array}{cc}
u_0 + 2\sqrt{\Delta}\sin2(\lambda-\lambda_{0})   &  \varepsilon =1\\ \displaystyle
 u_0+ A_0 e^{2\lambda}+\frac{\Delta}{A_0} e^{-2\lambda}  & \varepsilon =-1 \\ \displaystyle
 (E^2-L^2)(\lambda-\lambda_{0})^2 +\frac{J E L-M L^2}{E^2-L^2}  & \varepsilon =0 , E^2\neq L^2 \\
\lambda_{0}+ 2\sqrt{L^2(M\mp J)}\lambda  &\varepsilon =0 , E=\pm L,
\end{array} \right.
\end{equation}
where $\lambda_0, A_0$ are arbitrary constants and
\begin{align} \label{radius-delta}
u_0 = \frac{M}{2} +\frac{E^2 - L^2 }{2\varepsilon}, \quad
\Delta=\frac{\left(M+J+\varepsilon(E+L)^2\right) \left(M-J+\varepsilon(E-L)^2\right)}{16}.
\end{align}

Equations (\ref{dott}) to (\ref{dotu}) are the same for all BTZ spacetimes \cite{Cruz1994, Martinez:2019nor}. Although the radial motion is easily determined for each BTZ geometry, the behavior of $t(\lambda)$ and $\theta(\lambda)$ is strongly dependent on the type of zeros of $R(u)$ in the  denominators of \eqref{dott} and \eqref{dottheta}. Table \ref{tablerootsR(u)} shows the roots of $R(u)$ for the different types of the BTZ spacetimes as functions of $M$ and $J$. Note that $R(u)$ has positive roots only for black holes, while $R(u)$ is a positive-definite function for $u\in(0,\infty)$, with negative or complex roots in other geometries. 

For all BTZ spacetimes, timelike geodesics require $\Delta \ge 0$. Timelike circular geodesics of radius $\sqrt{u_0}$ appear under the conditions $u_0>0$ and $\Delta=0$. Spacelike circular geodesics are obtained in the limit $\Delta \to 0$, with $A_0= \text{constant}\times \sqrt{|\Delta|}$. Null circular geodesics only exist for $E=\pm L$ in geometries with $M=\pm J$ and can have any radius, as discussed in section \ref{circularorbits}.

\begin{table}[ht]
\centering
\caption{Roots of $R(u)$ for different  values of $M$ and $J$, and the corresponding BTZ geometry. }
\begin{tabular}{c|c|c}
\hline 
\bf{ $\boldsymbol{M}$ - $\boldsymbol{J}$ region}  & \bf{Roots of $\boldsymbol{R(u)}$ } &  \bf{Geometry} \\
\hline \hline 
$M > 0$ and $|J| < M $ & $ \frac{1}{2} \left(M \pm\sqrt{M^2-J^2}\right) >0$& Generic black hole \\ \hline  
$M > 0$ and  $|J| = M $ & $ \frac{1}{2}M >0 $  & Extremal black hole \\  \hline 
$M < 0$ and $|J| < -M $ & $\frac{1}{2} \left(M \pm\sqrt{M^2-J^2}\right) <0$& Naked conical singularity\\  \hline 
$M < 0$ and $|J| = -M$ &  $ \frac{1}{2}M <0$ & Extremal naked singularity \\  \hline 
$M = 0$ and $J=0 $ & $0$ & Massless BTZ geometry \\  \hline 
$M = -1$ and $J=0 $ & $0, -1$& AdS$_3$ vacuum \\  \hline 
$J^2>M^2$ & $\frac{1}{2} \left(M \pm i \sqrt{J^2-M^2}\right) \in \mathbb{C}$&Overspinning naked singularity\\ 
\hline 
\end{tabular}
\label{tablerootsR(u)}
\end{table}

\subsubsection{Radial bounds} 
Non-circular geodesics in the BTZ geometries can be bounded, but they can also reach infinity or fall into the central singularity. In order to examine these bounds on the radial motion, it is convenient to write \eqref{dotu} as
\begin{align}
    \label{hu}
    \frac{\dot{u}^2}{4}=h(u)=-\varepsilon u^2 +B u -C, 
\end{align}
where $B$ and $C$ depend on the geometry ($M,J$) and the constants of integration ($E,L$),
\begin{align}
B=\varepsilon M+ E^2 - L^2\quad \text{and} \quad C=JEL-M L^2+\frac{\varepsilon J^2}{4}.
\end{align}

Table \ref{table-radial-bounds} contains all possible geodesics that exist in the region $h(u) \geq 0$ for $u > 0$. For null geodesics, $h(u)$ is just a linear function and then it has only one root provided\footnote{The case $E^2=L^2$ leads to null circular geodesics, which are discussed in Sec. \ref{circularorbits} } $E^2\neq L^2$. If that root is positive, it would define an upper or lower bound for $u$, otherwise the null geodesic extends from $0<u<\infty$. Thus, three types of non-circular null geodesic are possible.  For non-null geodesics, $h(u)$ is a quadratic function whose positive roots $u_{\pm}$ (with $u_{+}\geq u_{-}$) are the bounds for the radial motion. Timelike  geodesics do not reach infinity and only two types exist. Spacelike geodesics appear in three different types. In two of them, the geodesics  extend between infinity and a lowest bound, or from infinity to the central singularity. In the remaining  type, the spacelike geodesic goes from the singularity to a finite upper bound. A general feature follows from \eqref{hu}:  since $\dot{u}^2(u=0)= -4C$, then $C<0$ is a necessary condition for geodesics reaching $u=0$. The previous analysis amounts eight types of noncircular geodesics with different radial bounds depending on the causal type, given by $\varepsilon$, and the values of $B$, $C$ and the discriminant $\Delta$.  These different geodesics are separately analyzed in Sec. \ref{GONS} for the overspinning spacetime.

\begin{table}[ht!] 
\centering
\caption{Radial bounds in terms of the constants $B=\varepsilon M+ E^2 - L^2, C=-L^2 M +JEL+\varepsilon J^2/4, \Delta=(B^2-4\varepsilon C)/16$ and $A_0$. Eight types of geodesics are found and labeled in the fourth column. Here $u_{\pm}$ are the positive roots of the equation $h(u)=0$ chosen as $u_{+}\geq u_{-}$.  }
\begin{tabular}{|c|c|c|c|}
\hline 
$\varepsilon$ & $B, C, \Delta$ and $A_0$ & Range of $u$ & Type \\[1.5mm] \hline 
 & $B > 0$ and $C > 0$ & $\frac{JEL-L^2 M}{E^2-L^2}\leq u < \infty$ & N$_1$ \\[1.5mm] \cline{2-2} \cline{3-3}   \cline{4-4} 
0 & $B \geq 0$ and $C \leq 0$ & $0\leq u <\infty$ & N$_2$ \\[1.5mm] \cline{2-2} \cline{3-3} \cline{4-4}  
&  $B<0$ and $C < 0$ & $0\leq u \leq \frac{JEL - L^2 M}{E^2-L^2} $ & N$_3$ \\[1.5mm]  \hline 
1 &  $C>0$, $B>0$, $\Delta>0$ & $u_{-} \leq u \leq u_{+}$ & T$_1$ \\[1.5mm] \cline{2-2} \cline{3-3}  \cline{4-4} 
 & $C \leq 0$ & $0 \leq u \leq u_{+}$ & T$_2$ \\[1.5mm] \hline 
 &\begin{tabular}{c} $B\geq 0$ and  $C \leq 0$ \\ $B\leq 0$ and  $\Delta \leq 0$ \end{tabular}& $0\leq u < \infty$  & S$_1$ \\[1.5mm] \cline{2-2} \cline{3-3}  \cline{4-4}
-1 & $B<0$, $C < 0$, $\Delta \ge 0$, $A_0 <0$ & $0\leq u \leq u_{-}$ & S$_2$ \\[1.5mm] \cline{2-2} \cline{3-3}  \cline{4-4}
 & \begin{tabular}{c}  $B<0$, $C < 0$, $\Delta \ge 0$, $A_0 >0$\\ $C > 0$ \end{tabular} & $u_{+}\leq u <\infty$   & S$_3$ \\[1.5mm] \hline 
\end{tabular} \label{table-radial-bounds}
\end{table}

\subsubsection{Behavior of the time coordinate} 
An important question is whether the time coordinate is monotonic along the geodesics. Assuming $\dot{t}(\lambda_1)>0$, is it possible that $\dot{t}(\lambda_2)<0$ for some $\lambda_2 > \lambda_1$? For this to happen with $t(\lambda)$ a continuous function, there must exist $\lambda_0$ for which $\dot{t}(\lambda_0)=0$, with $\lambda_1<\lambda_0< \lambda_2$. According to \eqref{dott} and $E\neq 0$, $\dot{t}$ vanishes for $u^*=JL/(2E)$. In order for this to be physically possible, the condition $JLE >0$ must be satisfied.
Moreover, for $\dot{t}$ to change sign at $u^*$ it is necessary that $\dot{u}|_{u^*}\neq 0$ ($\dot{u}|_{u^*}\in \mathbb{R}$) and therefore $\dot{u}^2|_{u^*}=4 h(u^{*})>0$. From \eqref{dotu}, this condition means
\begin{align}\label{sign_tdot}
    h(u^{*})=-\left(\frac{L^2+ \varepsilon u^* }{u^*}\right)\left((u^*)^2 - M u^* +\frac{J^2}{4}\right)>0.
\end{align}
Since $u^*>0$, the first factor in the above expression is always positive for timelike and null geodesics ($\varepsilon \geq 0$). The second factor is $R(u^*)$, which is nonnegative for all values of $r>0$ in geometries with naked singularities and also for the extremal black hole. Hence, $h(u^{*})$ is nonpositive in these geometries and the sign of $\dot{t}$ never changes for timelike or null geodesics. This is consistent with the absence of closed timelike curves in the conical and overspinning spacetimes.

The only remaining case is the nonextremal rotating black hole\footnote{Consider the caveat that $t$ is not a well-defined coordinate across the horizons of the black hole spacetime.} ($M>|J|$), for which $R(u)$ is negative between the horizons, $r_{-} < r < r_{+}$. This means that $\dot{t}$ can change sign only for 
\begin{equation} \label{tchange}
r^2_{-} < u^* < r^2_{+},
\end{equation}
which is not surprising since $t$ is a spacelike coordinate in this region. Then, from \eqref{tchange}, a necessary condition for $\dot{t}$ to change sign is
\begin{equation} \label{ineq}
\frac{M}{|J|}-\sqrt{\left(\frac{M}{J}\right)^2-1} \; <\;  \text{sgn}(J)\frac{L}{E} \; < \;  \frac{M}{|J|}+\sqrt{\left(\frac{M}{J}\right)^2-1}.
\end{equation}
Since  $1 <(M/|J|)<\infty$ holds for nonextremal rotating black holes, the inequality \eqref{ineq} can always be obeyed in some range of values of $L/E$.

For spacelike geodesics ($\varepsilon=-1$) the first factor in \eqref{sign_tdot} reduces to $2 E L/J-1$, which has no definite sign. Hence, there are spacelike geodesics in the BTZ geometries where the sign of $\dot{t}$ changes.

\subsubsection{Rotational dragging}  
\begin{figure}[h!]
    \centering
    \includegraphics[scale=0.18]{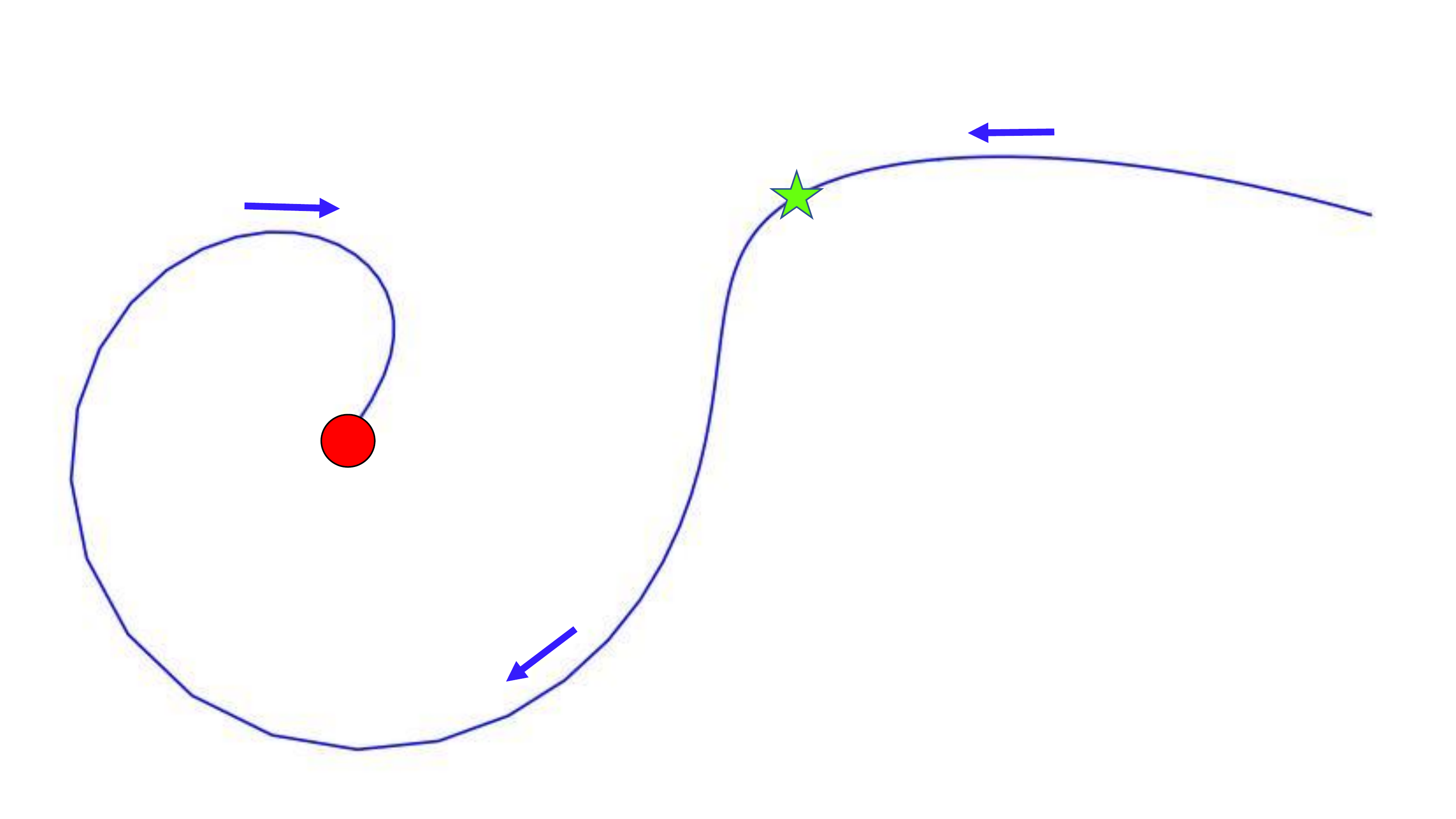}
    \caption{Null geodesic of type N$_2$, with $E>0$ and  $JL<0$, dragged into the singularity (the circle in red at $u=0$). The massless particle approaches the singularity from infinity as the arrows indicate. Using the usual convention for an increasing angle in polar coordinates, the figure shows that far from the singularity $\dot{\theta}$ is positive until the point indicated with a green star ($\Bar{u}$ given by \eqref{ubar}) where vanishes. After that, $\dot{\theta}$ becomes negative and the massless particle is dragged towards the singularity.}
    \label{drag}
\end{figure}

Since $R(u)$ has a definite sign, as explained in the previous section, it  follows from \eqref{dottheta} that $\dot{\theta}$ could change sign at a radius given by 
\begin{align} \label{ubar}
\Bar{u}=M-JE/(2L)>0.
\end{align}
This requires $L\neq0$. Moreover, for $J=0$, \eqref{dottheta} reduces to $\dot{\theta}=L/u$, which never changes sign. Thus, $J L \neq 0$ is a necessary condition for a sign change in $\dot{\theta}$. Similarly to the previous analysis for $\dot{t}$, the condition $h(\Bar{u})>0$ is also required. This necessary restriction is
\begin{align}\label{rot_dragging}
    h(\Bar{u})=-\left(\frac{2 E L}{J}+\varepsilon\right) R(\Bar{u}) >0.
\end{align}
Thus, in the case of causal geodesics ($\varepsilon \ge 0$), the  requirement 
\begin{align}\label{rot_drag2}
    2 E L/J+\varepsilon<0,
\end{align}
appears as a necessary condition for sign reversal of $\dot{\theta}$ for BTZ geometries where $R(u)>0$.\footnote{For $J\neq0$, $R(u)<0$ only in the region between the horizons of nonextremal black holes. Then, the necessary condition has the opposite sign, $2 E L/J+\varepsilon>0$.} Condition \eqref{rot_drag2} additionally implies  $JEL<0$. For a large radius $\dot{\theta}$ has the sign of $L$, and considering $E>0$, a change in the sign of $\dot{\theta}$ is only possible if $J$ has the other sign. Near $u=0$, $\text{sgn}(\dot{\theta})= \text{sgn}(J) = -\text{sgn}(L)$. In this way, under the conditions \eqref{ubar}, \eqref{rot_drag2}, and $JEL<0$, a causal geodesic is dragged by the rotation of geometry as shown in Fig. \ref{drag}. 

For spacelike geodesics, the expression $2 E L/J+\varepsilon>0$ does not a definite sign. Hence, these geodesics could be also dragged.

\subsection{Circular geodesics} \label{circularorbits} 
A geodesic is said to be circular if its radius is constant. The radius $r_0=(u_0)^{1/2}\ell \neq 0$ is constant provided $\dot{u}=0$ and $\Ddot{u}=0$ at $u_0$. From \eqref{dotu} and \eqref{u_dotdot}, these conditions correspond to
\begin{align} \label{circularcond1}
0=& -\varepsilon \left(u_0^2 -M u_0 + \frac{J^2}{4}\right) + (E^2-L^2)u_0 + M L^2  - JEL, \quad \text{and} \\ \label{circularcond2}
0=& - 2\varepsilon\, u_0+\varepsilon M + E^2 - L^2.
\end{align} 

\begin{table}[h!] 
\resizebox{0.80\textwidth}{!}{\begin{minipage}{\textwidth}
\captionsetup{margin={0cm,-4cm}}
\caption{ Existence of circular geodesics in BTZ geometries.  The conditions on the constants of motion $E$ and $L$ are given for different regions in the $M-J$ plane. The parameter $\alpha$ is defined as $\alpha:=L/E$, and $\alpha_{\pm}$ denotes the roots of Eq. \eqref{discriminant} as displayed in \eqref{eqL2}. There are additional time and spacelike circular geodesics in two special cases ($L=0$ and $E=0$) not displayed in this table. See \eqref{L=0} for $L=0$ and \eqref{E=0} for $E=0$. Null circular geodesic are present only in the extreme cases $J=\pm M$ with $L=\pm E$. } \label{circular_options}
\begin{tabular}{c|c|c} 
\hline 
\bf{Condition on $M, J$} & \bf{Timelike}  &  \bf{Spacelike}\\[1.5mm] \hline \hline 
$J^2< M^2$, $M>0$ & \bf{---}  &  $  \begin{array}{c}\alpha_{+} \in (-\infty,\frac{J}{2M -J})\cup[1,+\infty) \;\text{or} \\ \alpha_{-} \in (-\infty,-1]\cup(\frac{J}{2M +J},+\infty) \end{array}$ \\[1.5mm] \hline 
 $J^2< M^2$, $M<0$ & $\alpha_{+} \in (\frac{J}{2M -J},1)$ or $\alpha_{-} \in (-1,\frac{J}{2M +J})$   &  \bf{---}\\[1.5mm] \hline \hline 
  $J^2 > M^2$, $M>0$ & $L=\pm E$  &  \bf{---}\\[1.5mm] \hline
 $J^2> M^2$, $J>0$ & $J >2M$, $\alpha \in (\frac{J}{2M -J},1)$  & $J>-2M$, $\alpha \in (-\infty,-1)\cup(\frac{J}{2M +J},+\infty)$  \\  \cline{2-2} \cline{3-3}  
 & $J=2M $, $M>0$,  $\alpha<1$ &    $J=-2M $, $M<0$,  \, $ \alpha<-1$\\
  \cline{2-2} \cline{3-3} 
  & $J<2M, M>0$,  $\alpha \in (-\infty,1)\cup(\frac{J}{2M -J},+\infty)$  &    $J<-2M$, $M<0$, $\alpha \in (\frac{J}{2M +J},-1)$\\[1.5mm] \hline 
 $J^2> M^2$, $J<0$ & $J >-2M$,  $\alpha \in (-\infty,\frac{J}{2M +J})\cup(-1,\infty)$  & $J>2M$, $M<0$,  $\alpha \in (1,\frac{J}{2M -J})$\\ \cline{2-2} \cline{3-3}  
 & $J=-2M $, $M>0$,  $\alpha>-1$ &    $J=2M $, $M<0$,  \, $ \alpha>1$\\
  \cline{2-2} \cline{3-3} 
  & $J <-2M $, $\alpha \in (-1,\frac{J}{2M +J})$ &    $J <2M $, $\alpha \in (-\infty,\frac{J}{2M -J})\cup(1,\infty)$ \\[1.5mm] \hline \hline
 $J=\pm M$, $M>0 $ & $L=\pm E$ &   $\begin{array}{c}L=\pm E \\ -|L| < E < 3|L|, J L >0 \\   -3|L| < E < |L|, J L <0 \end{array} $\\[1.5mm] \hline 
 $J=\pm M$, $M<0 $ & $ \begin{array}{c} (E < -3|L|)\cup (E> |L|), J L >0 \\ (E < -|L|)\cup (E>3 |L|), J L <0 \end{array} $ &  \bf{---} \\[1.5mm] \hline 
\end{tabular} 
\end{minipage}}
\end{table}

For null geodesics ($\varepsilon=0$), conditions \eqref{circularcond1}-\eqref{circularcond2} reduce to $ML^2=JEL$ and $E^2 =L^2$. Therefore, null circular geodesics are possible only in extreme BTZ spacetimes with $M = \pm J$, including the massless case $M=J=0$. In these cases, the constants of motion are related as $E =\pm L$,  and the circular geodesic can have any radius, in agreement with the results in \cite{Cruz1994} and \cite{Martinez:2019nor}. Null circular geodesics are absent in overspinning spacetimes . 

For timelike and spacelike geodesics ($\varepsilon \neq 0$), using \eqref{circularcond2}, we obtain the general expression for the radius of a circular geodesic,
\begin{align} \label{radius}
\frac{r_0^2}{\ell^2} = u_0 = \frac{M}{2} +\frac{E^2 - L^2 }{2\varepsilon},
\end{align}
and the following relation linking the conserved charges $(M, J)$ and the constants of motion $(E, L)$: 
\begin{align} \label{discriminant}
(\varepsilon M + E^2 - L^2)^2 =4\varepsilon \left(J E L -M L^2 + \varepsilon \frac{J^2}{4}\right),
\end{align} 
which matches the condition $\Delta=0$  previously announced for non-null circular geodesics (cf. Eq.\eqref{radius-delta}). This can be understood as follows: $h(u)$ is a quadratic function of $u$ whose discriminant is $16 \Delta$. The requirements of a circular geodesic, $\dot{u}=\ddot{u}=0$,  are only satisfied if $h(u)= -\varepsilon(u-u_0)^2$, namely, if the square radius of the geodesic, $u_0$ is a double root of $h(u)=0$, which is equivalent to demanding $\Delta=0$.

For the circular geodesics with $L=\pm E$, Eqs. \eqref{radius} and \eqref{discriminant} lead to
\begin{align}\label{alphapm1}
u_0 = \frac{M}{2}  \quad \text{and} \quad  M^2- J^2 = -4\varepsilon E^2 (M \mp J).
\end{align}
The first condition above can only be satisfied in BTZ spacetimes of positive mass. The second equation leads to two scenarios. In the first, the equation is fulfilled by extreme BTZ black holes with $M = \pm J >0$ for any $\varepsilon$. The existence of timelike circular geodesics was announced in \cite{Cruz1994}, while the spacelike case seems to be original. A second scenario appears for non-extremal spacetimes ($M^2 \neq J^2$) where the second equation in \eqref{alphapm1} reduces to
\begin{align}\label{alpha2pm1}
 M\pm J = -4\varepsilon E^2.
\end{align}
Since $E\neq 0$ (there are no geodesics if $E$ and $L$ simultaneously vanish), the timelike case ($\varepsilon=1$) requires $J^2 > M^2$. Thus, these particular timelike circular geodesics are only present in the overspinning spacetime, while the spacelike ones also exist in the black hole geometry.

Before analyzing the general case $E^2 \neq L^2$, we proceed with the particular choices $L=0$, $E=0$, and $L\neq \pm E$ with $M=\pm J$.

For $L=0$, Eqs. \eqref{radius} and \eqref{discriminant} become
\begin{align}\label{L=0}
u_0 = \frac{M+\varepsilon E^2}{2}  \quad \text{and} \quad \left(M + \varepsilon E^2\right)^2 = J^2,
\end{align}  
respectively. Thus, $L=0$ timelike circular geodesics always exist in the overspinning spacetime if $ M\ge0$ and for a negative mass the condition $E^2+2 M >0$ is required. These circular geodesics exist in the rotating conical spacetimes provided $E^2+M>0$ and $E^2+2 M \le 0$, namely $-M<E^2 \le -2 M$, and they are absent in the BTZ black hole. The existence of spacelike circular geodesics with $L=0$ requires $M>0$. Thus, they are present in the rotating black hole spacetime provided $E^2<M$ and do not exist in the overspinning or conical spacetimes.

For $E=0$, Eqs. \eqref{radius} and \eqref{discriminant} yield
\begin{align}\label{E=0}
u_0 =\frac{M -\varepsilon L^2}{2}  \quad \text{and} \quad \left(M + \varepsilon L^2 \right)^2 = J^2,
\end{align}  
respectively. Therefore, the $E=0$ timelike circular geodesics in the overspinning spacetime need $M > L^2>0$, and do not exist in the black hole or conical spacetimes.  The spacelike case with $E=0$ are present in the black hole provided that $L^2 \le 2M$, and they are absent in the conical geometry. In the overspinning spacetime, the spacelike circular geodesics require $L^2 > 2M$ if $M \ge 0$ or $L^2 > -M $ if $M < 0$. 

Now, we consider the non-null geodesics with $L\neq \pm E$ in the extreme spacetimes $M =\pm J$.  Eq. \eqref{discriminant} gives
\begin{align} \label{extremes}
\left( E \pm L \right)^2 = - 2\varepsilon M,
\end{align} 
which means that $\varepsilon M <0$ is required for this class of circular geodesics in the extreme spacetimes. Then, these geodesics are spacelike for the extreme black hole and timelike for the extreme conical singularity. Using the mass given in \eqref{extremes}, the radius of the geodesic is
\begin{align}\label{extremeradius}
u_0 = \frac{\varepsilon}{4}(E\pm L)(E\mp 3L).
\end{align}
Therefore, the spacelike circular geodesic of the extreme black hole is defined in the range $(E\pm L) (E\mp 3L)<0$ and the range for the timelike circular case of the extreme naked singularity is  given by $(E\pm L)(E\mp 3L) >0$.

In order to handle \eqref{discriminant} in the general case $L^2 \neq E^2$  and $J^2 \neq M^2$ we introduce $\alpha:=L /E $ if $E\neq0$. After some algebra, Eq. \eqref{discriminant} turns out to be quadratic in $E^2$, whose two roots, specified with the sign $\pm$,  take the simple form
\begin{align} \label{eqL2}
E^2= \frac{\varepsilon(\pm J - M)}{(1\mp \alpha)^2},
\end{align} 
with $\alpha \neq \pm 1$. The existence of real solutions of \eqref{eqL2} is given by the condition $\varepsilon(\pm J - M)>0$, which discards timelike circular geodesics in the non-extremal black hole and spacelike ones in the non-extremal conical spacetime. Note that \eqref{eqL2} is a necessary condition for circular geodesics, but it is not sufficient to ensure their existence. The radius $r_0^2$ in \eqref{radius} must be also positive. Replacing \eqref{eqL2} in \eqref{radius} we obtain
\begin{equation} \label{r_0}
u_0 = \frac{J(\alpha \pm 1) \mp 2 \alpha M}{2 (1\mp\alpha)} = \frac{\pm J(1-\alpha^2)\mp 2 M \alpha(1\mp\alpha)}{2 (1\mp\alpha)^2}.
\end{equation}
Therefore, non-extremal spacetimes contain non-null circular geodesics (in the general case $L^2 \neq E^2, E \neq 0, L \neq 0$) under the conditions
\begin{equation} \label{cond-gen}
\varepsilon (\pm J - M)>0  \quad \text{and} \quad \pm J(1-\alpha^2) \mp 2
M \alpha(1\mp\alpha)>0.
\end{equation}
The ranges of the constants of motion set by the above conditions are summarized in Table \ref{circular_options} for the different BTZ geometries. The solutions $\alpha_{\pm}$ depend on the $\pm$ sign appearing in \eqref{eqL2}, therefore one solution can be obtained from the other by changing $J \rightarrow -J$ and $\alpha \rightarrow -\alpha$. 

The non-null circular geodesics presented in Table \ref{circular_options} are new results for the BTZ geometries, except for the timelike circular case of the extreme black hole ( $J=\pm M, M>0$), which was already reported in \cite{Cruz1994}.  The null circular geodesics for the extreme solutions ($M^2=J^2$) were previously found in the extreme black hole   \cite{Farina:1993xw,Cruz1994} and conical singularity \cite{Martinez:2019nor}. Additionally, the massless BTZ spacetime $M=J=0$ also contains null circular geodesics with arbitrary radius as shown in \cite{Martinez:2019nor}.

\section{Geodesics around an overspinning naked singularity } \label{GONS}
We now turn to the analysis of the geodesics in the overspinning spacetime, characterized by the condition $|J|>|M|$. As in the other BTZ geometries \cite{Cruz1994, Martinez:2019nor}, timelike, null and spacelike geodesics have specific features and extend over different regions of the entire geometry.

\subsection{Radial bounds } 

The radial bounds are equivalently determined by equations
\eqref{dotu} or \eqref{ulambda} imposing $\dot{u}^2>0$ or $u>0$, respectively. According to the causal parameter $\varepsilon$ and radial bounds, eight types of noncircular geodesics appear in the BTZ geometries, as shown in Table \ref{table-radial-bounds}. Appendix \ref{apendice} contains a detailed discussion of the radial bounds for causal and spacelike geodesics in the overspinning spacetime. In this appendix, each type of geodesic presented in Table \ref{table-radial-bounds} is related to specific conditions, mainly inequalities, depending on the constants of motion, $E$ and $L$, and the conserved charges $M$ and $J$ of the overspinning spacetime.  In what follows, we present a summary of Appendix \ref{apendice}. 

Three types of null geodesics are found in the overspinning spacetime: Geodesics reaching the infinity from a lower radial bound (type N$_1$), those extending from the central singularity to infinity (type N$_2$), and geodesics having a radial upper bound and that terminate at $r=0$ (type N$_3$).

As expected for an anti-de Sitter spacetime as the overspinning one, timelike geodesics cannot reach infinity. Thus,  they either loop around the singularity (type T$_1$) or fall into it (type T$_2$). A special case of timelike geodesics generated closed curves on the $r-\theta$ plane, which will be presented in Sec. \ref{autointer}.

Spacelike geodesics in the overspinning spacetime have similar bounds as the null geodesics and then can be classified in the three types, S$_1$ to S$_3$. However, as discussed in the Appendix \ref{apendice} (see also Eq. \eqref{ulambda}), the classification depends on a sign of a parameter fixed as initial condition. The two possible signs lead to two different types, S$_2$ and S$_3$.

\subsection{Projections on the $r$-$\theta$ plane} \label{orbits} 
We will now study the projections on the $r$-$\theta$ plane of geodesic lines. We will refer to them as the $r$-$\theta$ orbits. From \eqref{dottheta} and \eqref{dotu} we obtain $d\theta/du$ and then the integral 
\begin{equation} \label{intgen}
\theta-\theta_0=\int du\frac{u L-M L+JE/2}{2\left(u^2-M u+\frac{J^2}{4}\right) \sqrt{-\varepsilon u^2 + \left(E^2-L^2+M \varepsilon \right)u-J E  L-\frac{J^2 \varepsilon }{4}+M L^2 }},
\end{equation}
which provides the orbits for the BTZ geometries. In the cases where $J^2 \le M^2$ the quadratic expression, $u^2-M u+\frac{J^2}{4}$,  appearing in the denominator can be written as the product of two real linear functions of $u$. This factorization allows for a decomposition or \eqref{intgen} in partial fractions. However, this procedure is not  possible for the case  $J^2 > M^2$ since the quadratic does not allow such a factorization. A way to circumvent this obstacle is to consider the following substitution\footnote{See Section 2.25 in \cite{GR tables}.}
\begin{equation}
u=\frac{L (J E-M L)}{E^2-L^2}  +  \frac{J(E^2+L^2)-2 M E L }{2 \left(E^2-L^2\right)}  \frac{v-1}{v+1},
\end{equation}
leading to a more standard integral, which reduces to elementary functions as
\begin{align}
    \label{theta(r)}
    \theta-\theta_0 & =\frac{\sgn(J) }{4}\Bigg[ \frac{2}{\sqrt{|J|-M}}\text{arctan}\left( \frac{F(u; J,L,M, E)}{2\sqrt{|J|-M}} \right)\\ 
    \nonumber &+ \frac{1}{\sqrt{|J|+M}} \log  \left|\frac{2\sqrt{|J|+M}+F(u; -J,-L,M, E)}{2\sqrt{|J|+M}-F(u; -J,-L,M, E)}\right|  \Bigg],
\end{align}
where the function $F(u;J,L,M,E)$ is defined as 
\begin{align} \label{F}
 F(u; J,L,M,E):=\frac{- J(E+ L) +2u(E- L)+ 2LM}{\sqrt{-\varepsilon u^2 +(E^2-L^2+\varepsilon M)u+L^2 M -JEL-\varepsilon J^2 /4}}.
\end{align}

Equation \eqref{F} is valid for $J>0$. The expression for $J<0$ is obtained by replacing the pair $J,L$ by $-J,-L$ in \eqref{F}. This is consistent with the fact that this replacement changes the orientation of the angular coordinate $\theta$ in \eqref{intgen}, as expected.

For later convenience, it is useful to introduce the function
\begin{align} \label{su}
 s(u):=\sgn[- J(E+ L) +2u(E- L)+ 2LM].
\end{align} 
Since the expression in the denominator of Eq. \eqref{F} is just $\dot{u}$, $F(u;J,L,M,E)$ is a well-defined function in the domain of all the geodesics, except at the points where $\dot{u}=0$, namely, at the turning points or the case of circular orbits.  At those points, $F$ diverges but $\theta$ remains finite. 

\subsection{Self-intersections} \label{autointer} 
\begin{figure}[ht]
    \centering
    \begin{subfigure}{.30\textwidth}
        \centering
        \includegraphics[scale=0.27]{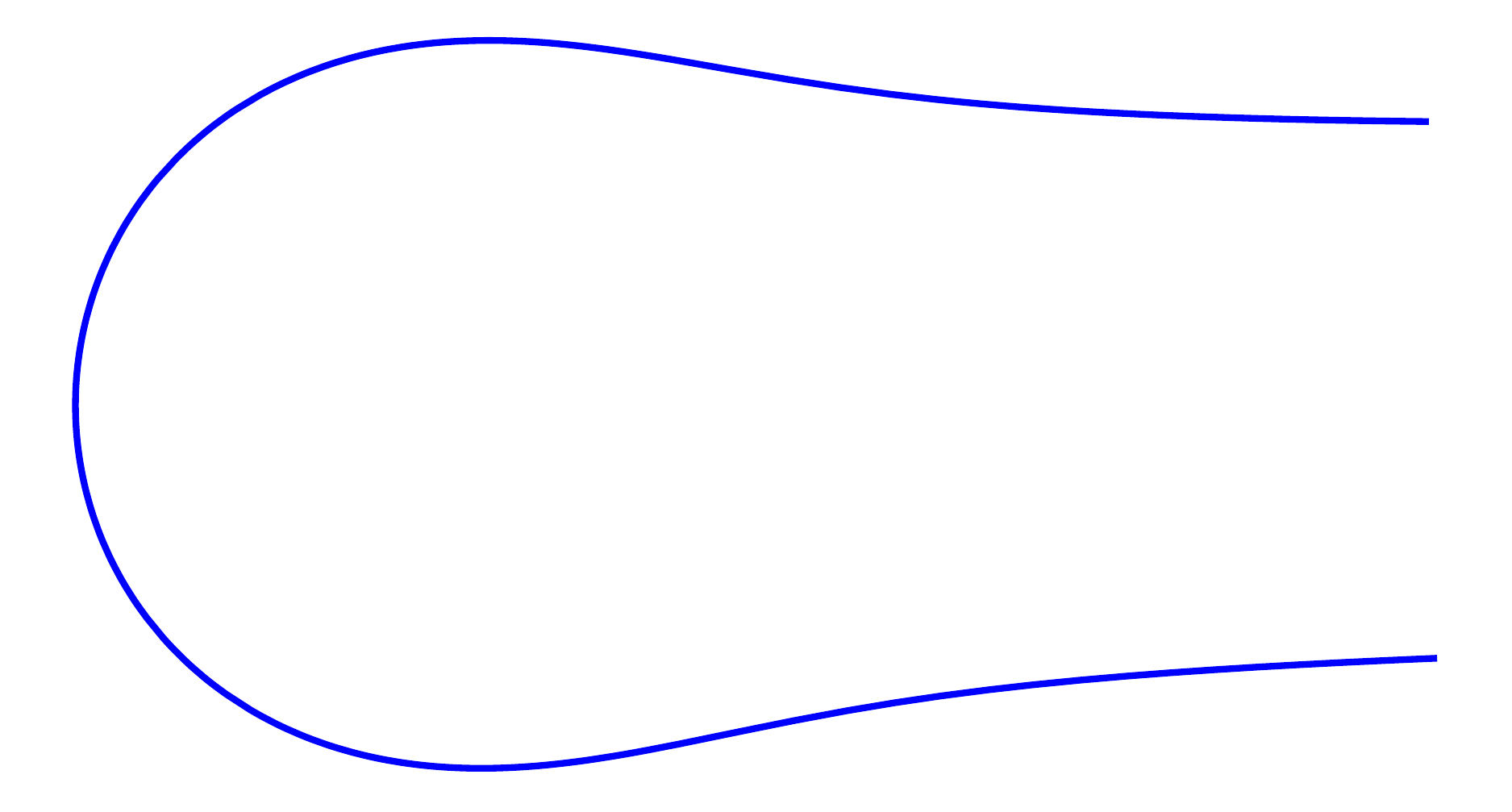}
        \caption{$N=0$} 
        \label{fig:my_label40}
    \end{subfigure}
    \begin{subfigure}{.30\textwidth}
        \centering
        \includegraphics[scale=0.2]{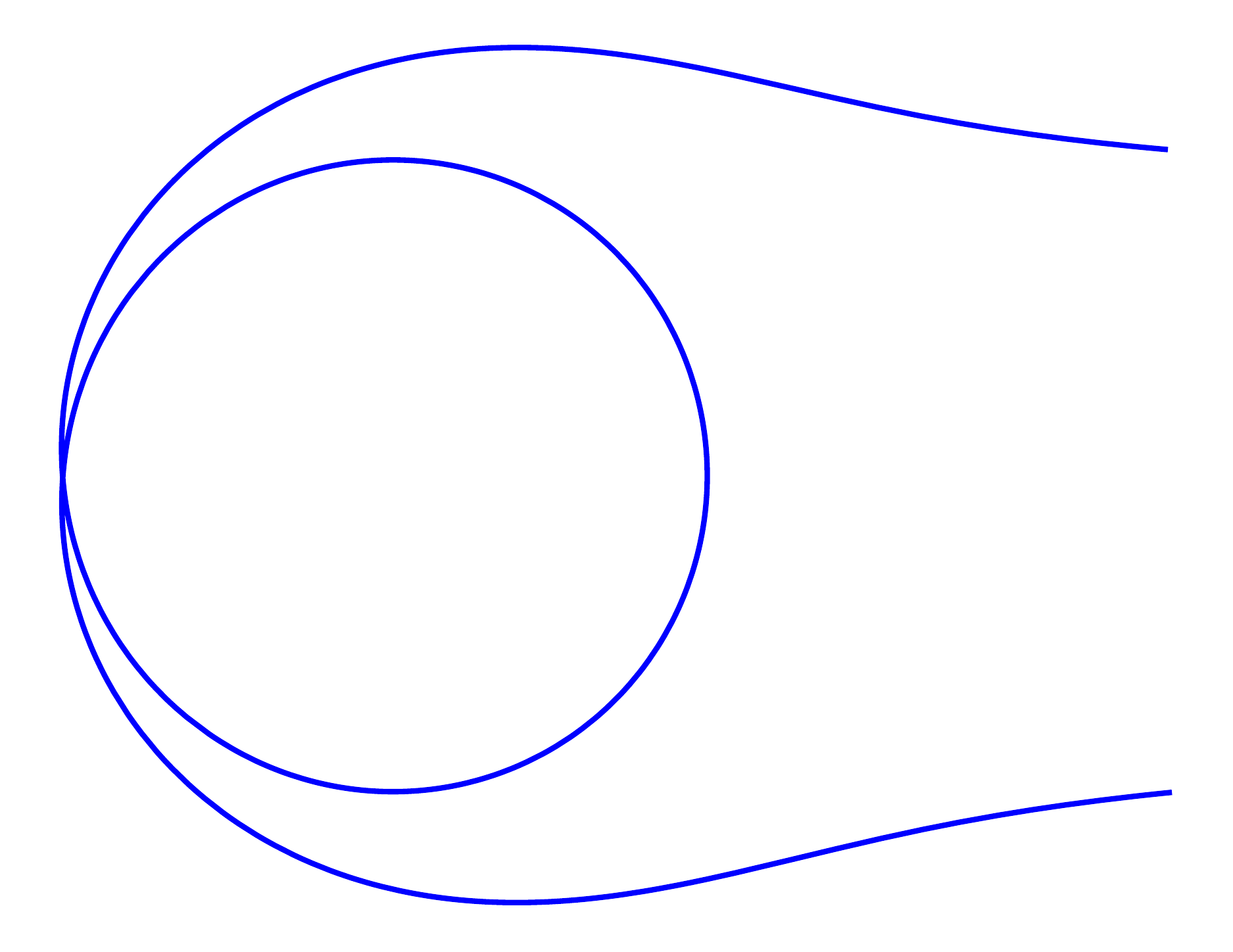}
        \caption{$N=1$} 
        \label{fig:my_label31}
    \end{subfigure}
    \begin{subfigure}{.30\textwidth}
        \centering
        \includegraphics[scale=0.24]{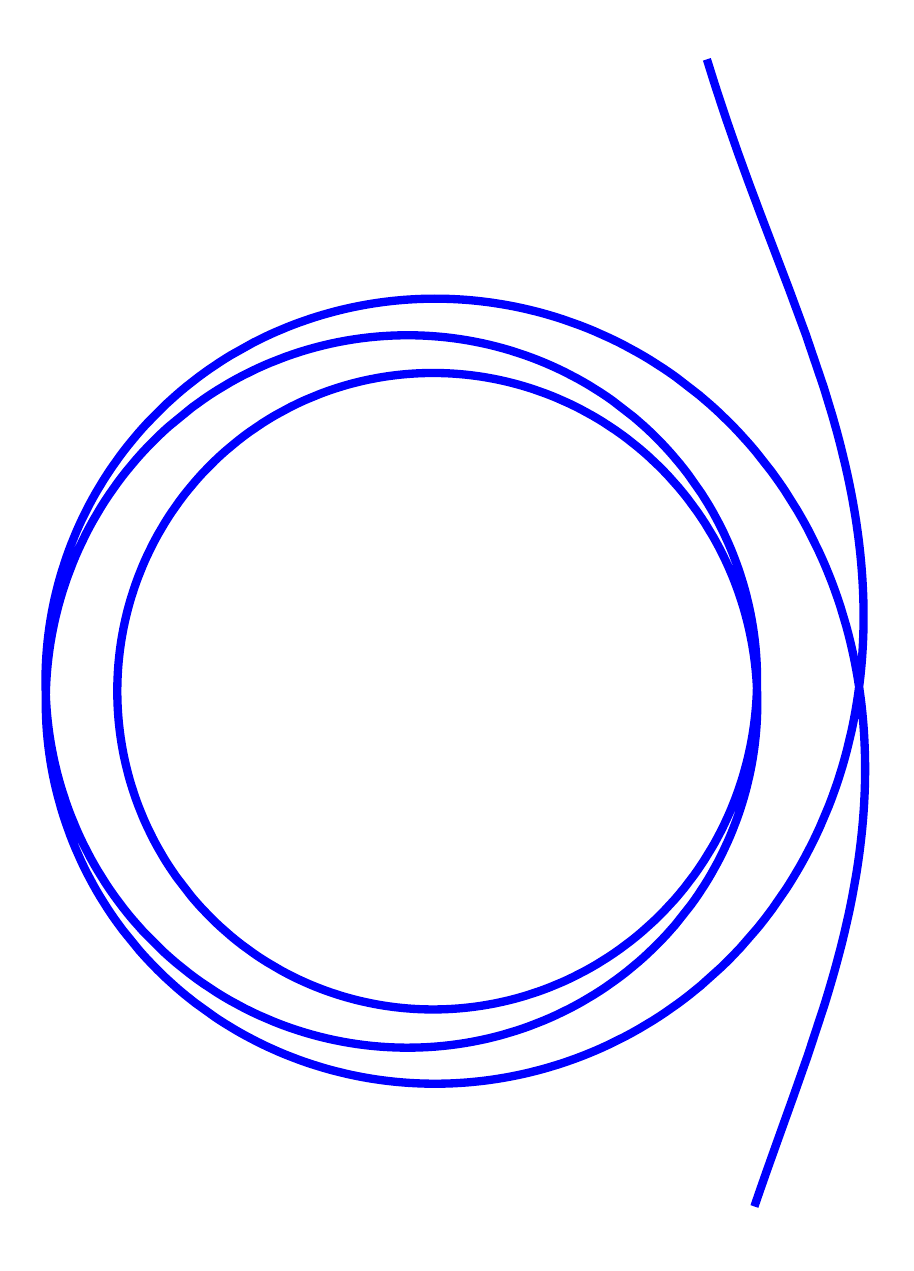}
        \caption{ $N=3$} 
        \label{fig:my_label32}
    \end{subfigure}
    \caption{Null geodesics of type N$_1$ with different number of self-intersections given by $N$.}
\label{null_plot2}
\end{figure}

\begin{figure}[h!]
    \centering
    \begin{subfigure}{.45\textwidth}
        \centering
        \includegraphics[scale=0.25]{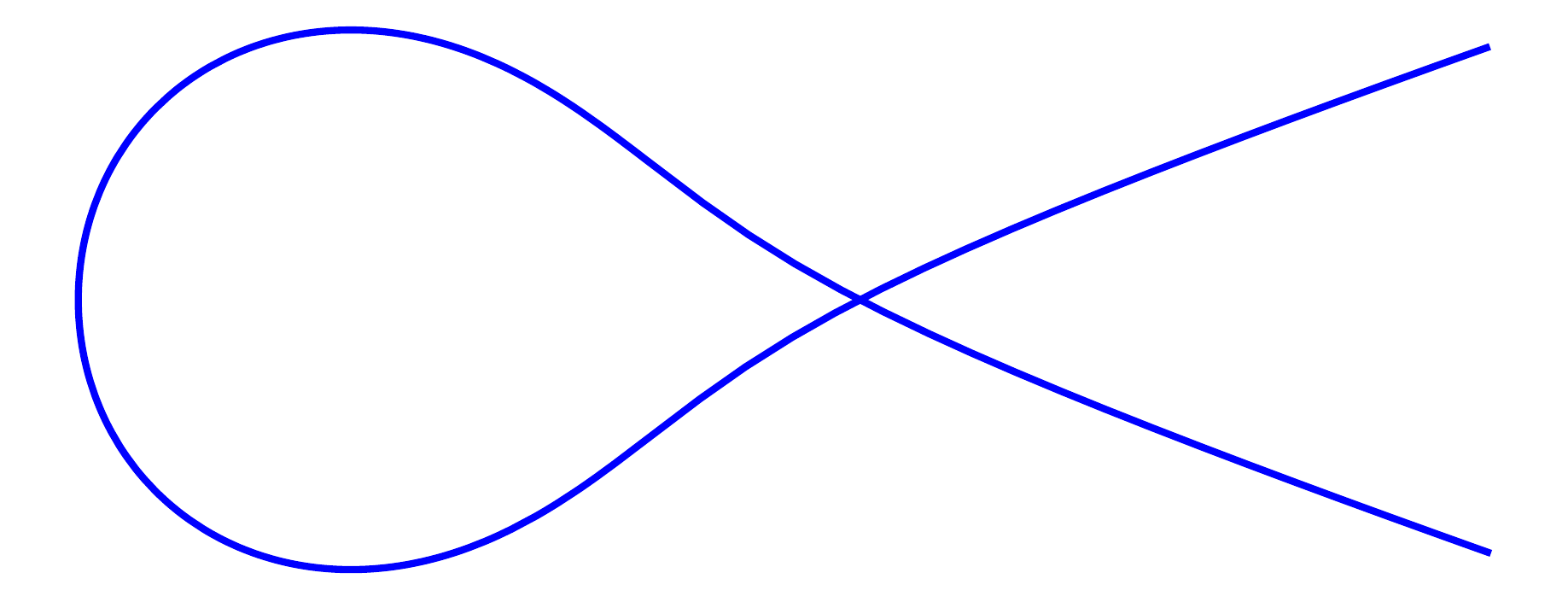}
        \caption{$N=1$} 
        \label{fig:my_label33}
    \end{subfigure}
    \begin{subfigure}{.45\textwidth}
        \centering
        \includegraphics[scale=0.25]{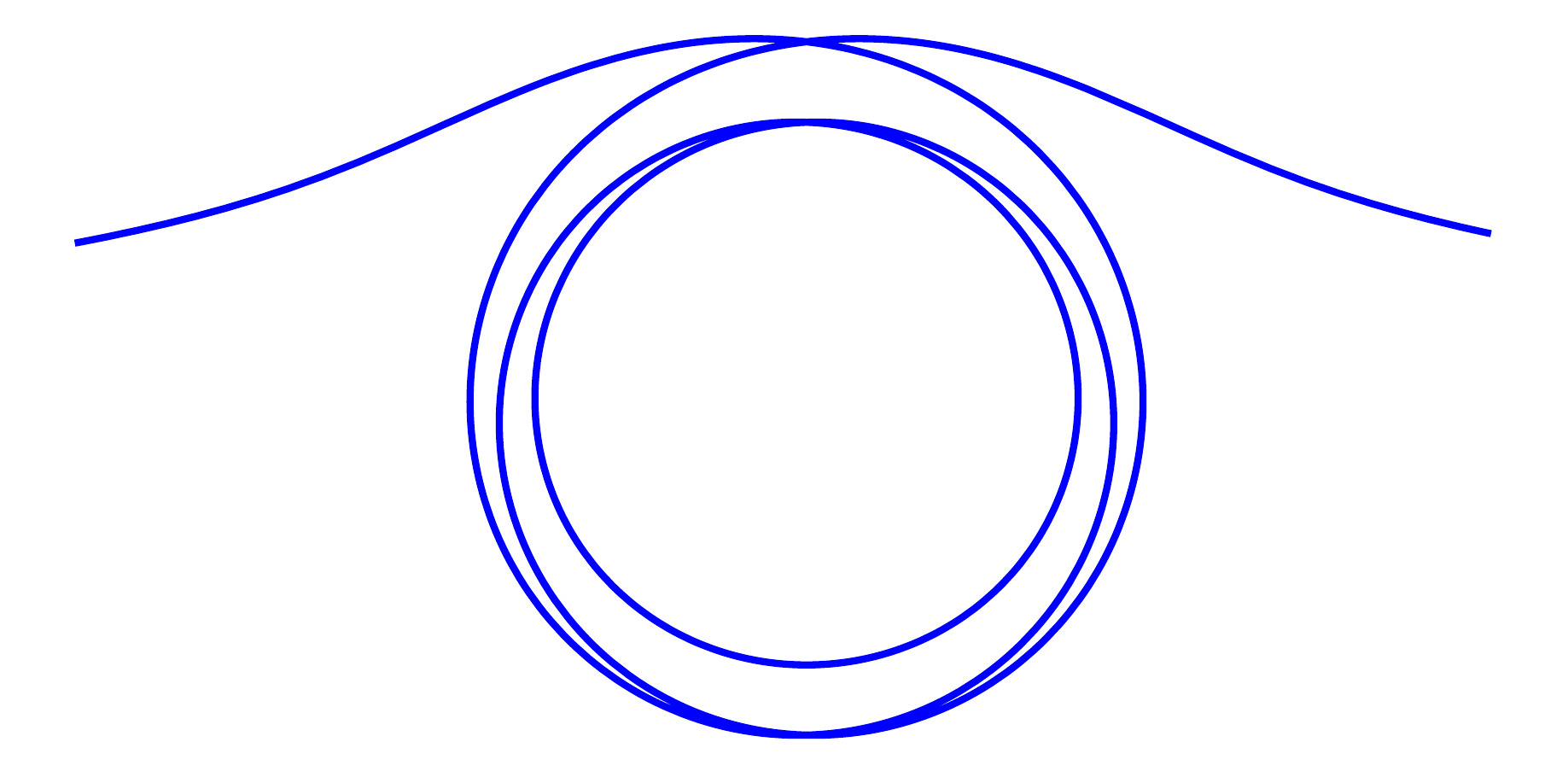}
        \caption{$N=3$} 
        \label{fig:my_label34}
    \end{subfigure}
    \caption{Spacelike orbits of type S$_3$ with $N=1$ and $N=3$ self-intersections. }
\label{SL_plot1}
\end{figure}

The orbits projected on the $r$-$\theta$ plane with a returning point can intersect themselves. There are two general cases. One in which the geodesics start at infinity and return to infinity, reach a non-vanishing minimum turning point for its radial motion, and go back to infinity. As shown in Table \ref{table-radial-bounds} (see also Appendix \ref{apendice}), types N$_1$ and S$_3$ behave in this way. The number of intersections can be obtained analyzing the ratio between the angle swept by $\theta$ from infinity to the turning point and $\pi$. This number counts how many times the geodesic turns around the origin, as shown in Figs. \ref{null_plot2} and \ref{SL_plot1}.

\begin{figure}[ht!]
    \centering
     \includegraphics[scale=0.3]{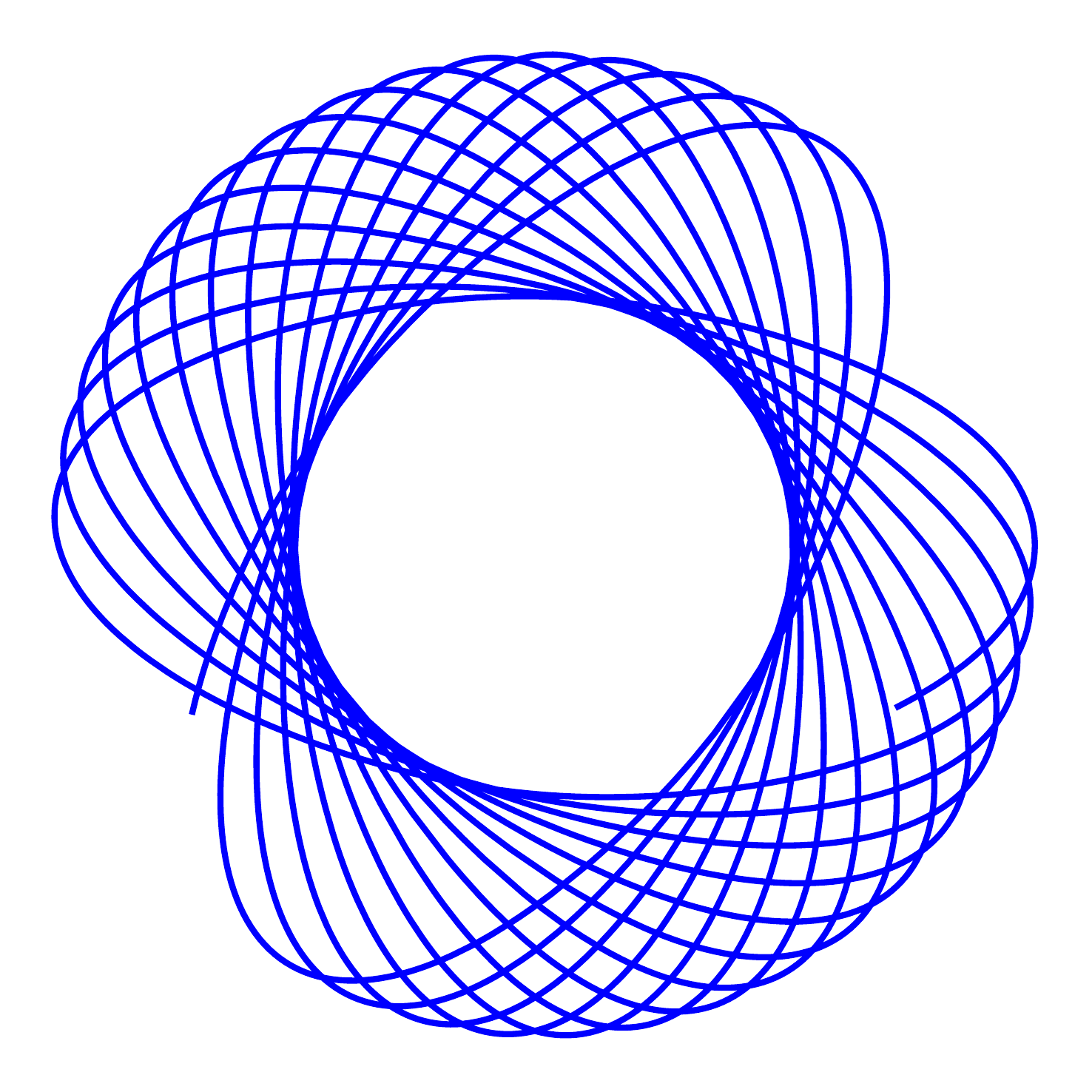}
    \caption{Timelike geodesic of type T$_1$. The orbits of these geodesics in the plane $r-\theta$ contain an infinity number of self-intersections. }
    \label{TL-several2}
\end{figure}

\begin{figure}[h!]
    \centering
    \begin{subfigure}{.30\textwidth}
        \centering
        \includegraphics[scale=0.15]{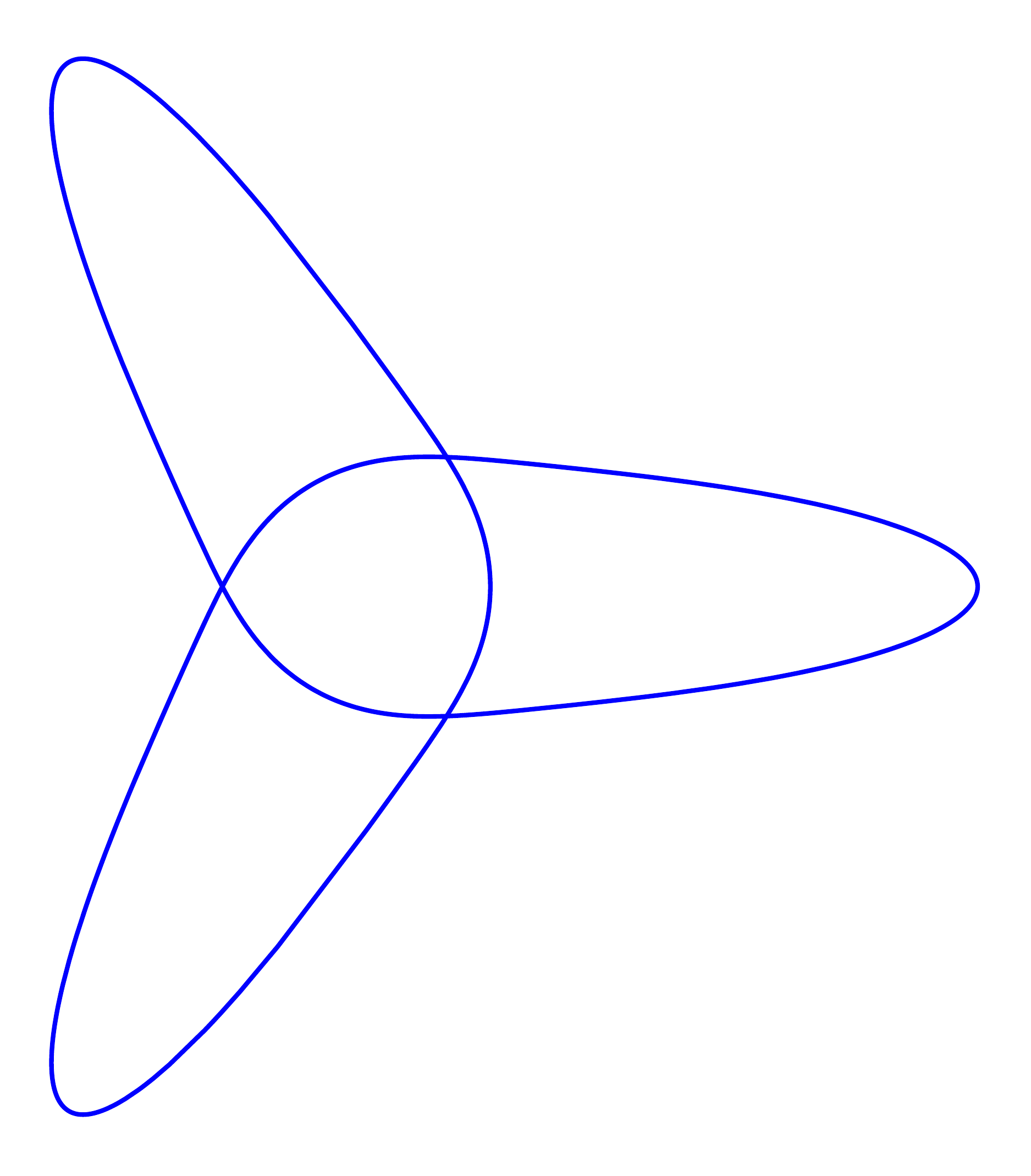}
        \caption{$\sqrt{|J|-M}=3/4$} 
        \label{fig:my_label21}
    \end{subfigure}
    \begin{subfigure}{.30\textwidth}
        \centering
        \includegraphics[scale=0.25]{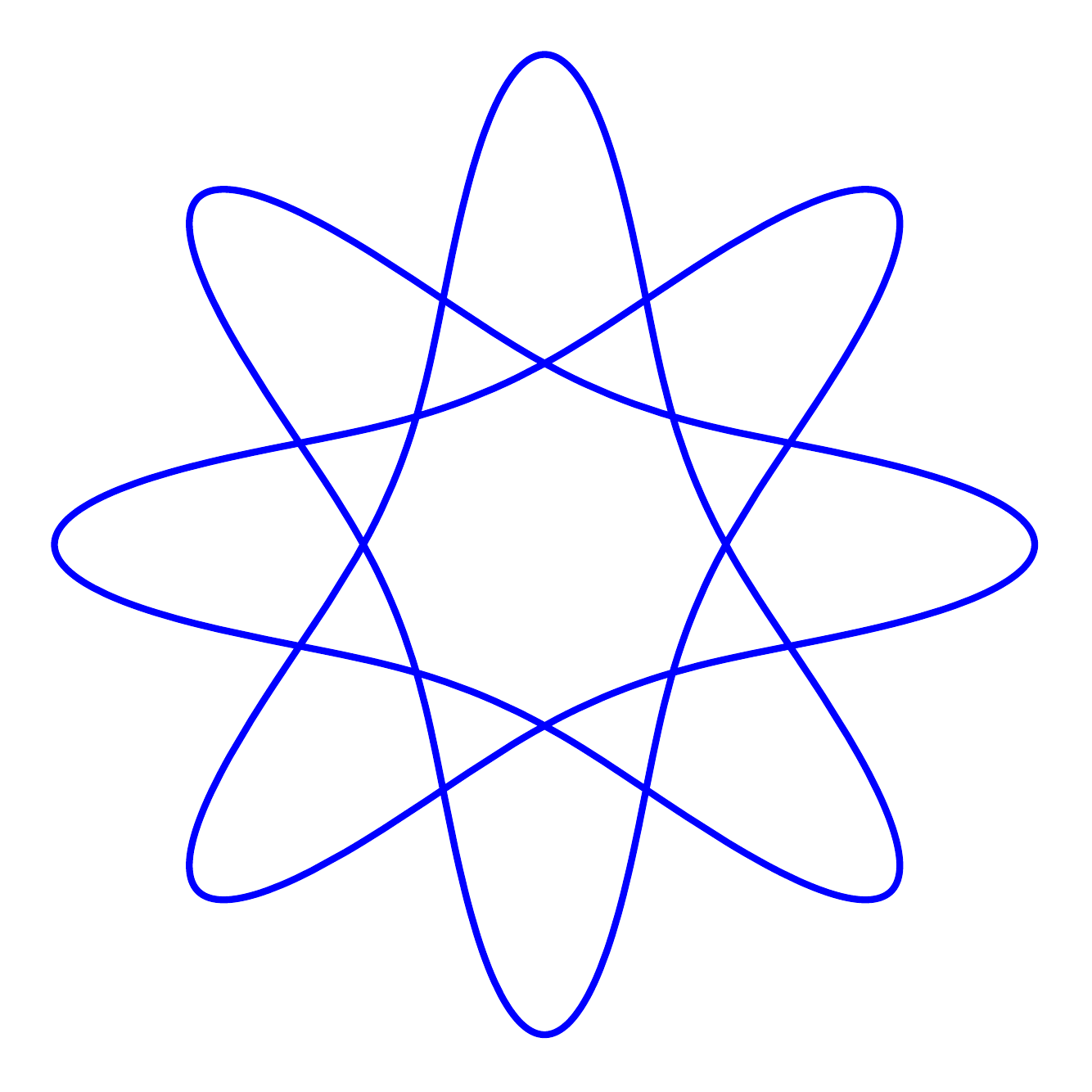}
        \caption{$\sqrt{|J|-M}=4/3$} 
        \label{fig:my_label22}
    \end{subfigure}
    \caption{Closed $r$-$\theta$ orbits of type T$_1$ with self-intersections where $\sqrt{|J|-M}$ takes rational values, where in (a) $|\Delta \theta|=\frac{2\pi}{3}$, in (b) $|\Delta \theta|=\frac{3\pi}{8}$.}
\label{TL_plot3}
\end{figure}

\begin{figure}[h!]
    \centering
    \begin{subfigure}{.30\textwidth}
        \centering
        \includegraphics[scale=0.25]{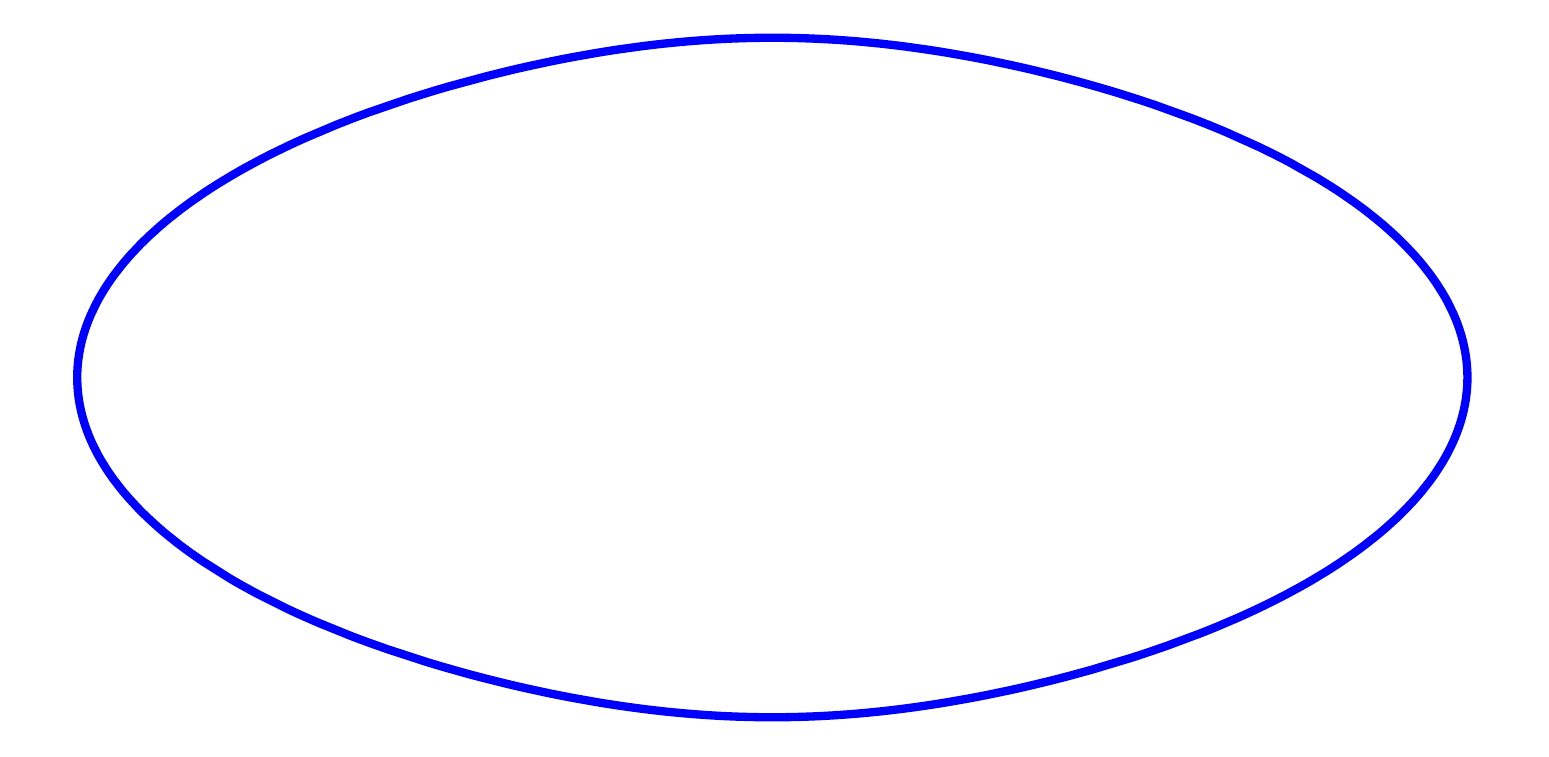}
        \caption{$\sqrt{|J|-M}=1$} 
        \label{fig:my_label10}
    \end{subfigure}
    \begin{subfigure}{.30\textwidth}
        \centering
        \includegraphics[scale=0.20]{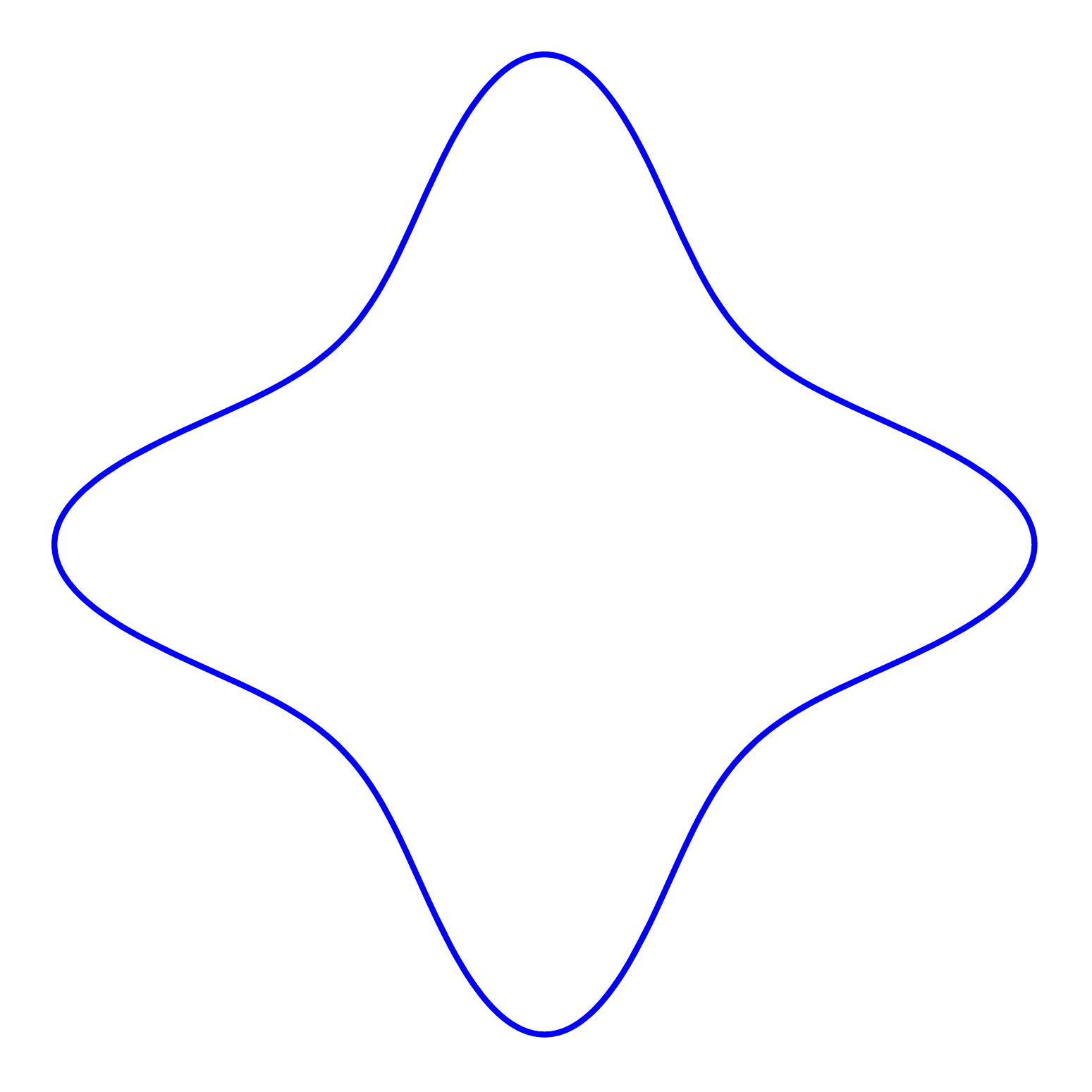}
         \caption{ $\sqrt{|J|-M}=2$. } 
        \label{fig:my_label11}
    \end{subfigure}
    \begin{subfigure}{.30\textwidth}
        \centering
        \includegraphics[scale=0.25]{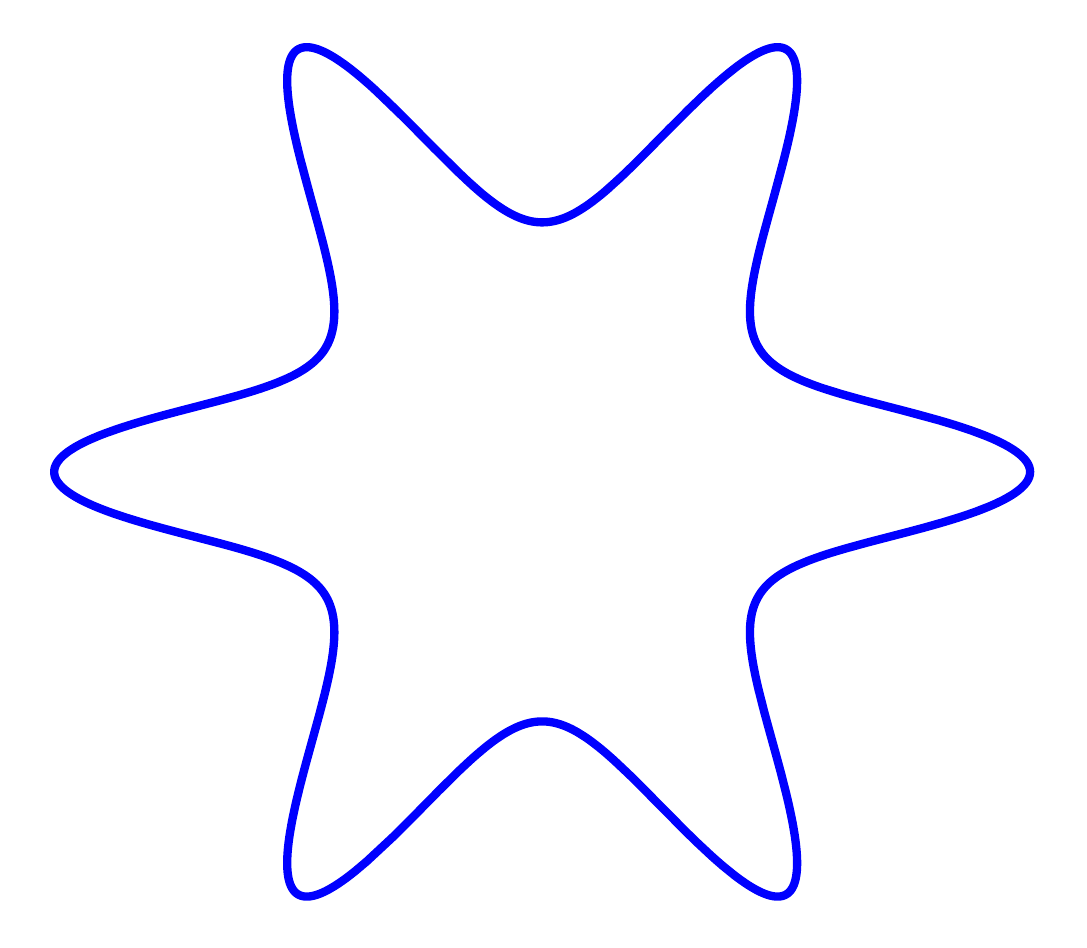}
        \caption{$\sqrt{|J|-M}=3$} 
        \label{fig:my_label12}
    \end{subfigure}
    \caption{Closed $r$-$\theta$ orbits of type T$_1$ geodesics without self-intersections. In the panels $\sqrt{|J|-M}$ is an integer, where in (a) $ |\Delta \theta|= \frac{\pi}{2}$, in (b) $ |\Delta \theta|= \frac{\pi}{4}$ and in (c) $ |\Delta \theta|= \frac{\pi}{6}$.}
\label{TL_plot2}
\end{figure}

In the second case, the radial motion is bounded by two finite radii given by the turning points. This is the case of type T$_1$ geodesics. In general, these orbits are not closed as their periastrons rotate and then they could have an infinite number of self-intersections (see Fig. \ref{TL-several2}).  However, if the values of $M$ and $J$ are finely tuned, the orbits can close with a finite number of self-intersections (see Fig. \ref{TL_plot3}), or without them (see Fig. \ref{TL_plot2}).

We start computing the number of self-intersections by analyzing geodesics with only one turning point. For the null geodesics of type N$_1$, the angle swept from infinity to the returning point $u_{\text{min}}=(JEL-L^2 M)/(E^2 - L^2)$, dubbed as $\Delta \theta$, is determined from \eqref{theta(r)} to be
\begin{align}
    |\Delta \theta| = \frac{\pi}{2 \sqrt{|J|-M}}.
\end{align}
Therefore, the number of self-intersections found in the N$_1$ null geodesics is
\begin{align} \label{N-null}
    N=\ceil*{\frac{|\Delta \theta|}{\pi}}-1=\ceil*{\frac{1}{2\sqrt{|J|-M}}}-1,
\end{align}
where the ceiling function $\ceil{x}$ gives the least integer greater than or equal to $x$. The ceiling function is necessary in \eqref{N-null} because if the value of $|\Delta \theta|/\pi$ is an integer, then there is one self intersection at infinity, so in this case the number of self intersections at a finite radius is actually $|\Delta \theta|/\pi-1$. Remarkably, the number of self-intersections does not depend on the geodesic parameters $E$ and $L$ and grows if $M>0$ approaches $|J|$, but it decreases when $-M>0$ is going to $|J|$. It is interesting to note that Eq. \eqref{N-null} is similar to the number of self-intersections found for null geodesics in the conical BTZ spacetime discussed in \cite{Martinez:2019nor}. Figure \ref{null_plot2}  shows three null geodesics with different number of self-intersections. 

In the case of S$_3$ spacelike geodesics, the angle swept from infinity to $u_{\text{min}}=u_+$ is given by 
\begin{align} \label{angSL}
    \Delta \theta = &\frac{1}{2\sqrt{|J|-M}} \text{arctan} \left( \frac{E- L}{\sqrt{|J|-M}} \right) + \frac{1}{4\sqrt{|J|+M}}\log \left| \frac{\sqrt{|J|+M}+(E+ L)}{\sqrt{|J|+M}-(E+ L)} \right| \nonumber \\ &+ \frac{\pi \,\sgn[J L]}{4\sqrt{|J|-M}},
\end{align}
where we have used the identity $s(u_{\text{min}})=-\sgn[J L]$, which is fulfilled by S$_3$ geodesics. The last term in \eqref{angSL} increases $\Delta \theta$ if $J$ and $L$ have the same sign. Fig. \ref{SL_plot1}
 shows two examples of spacelike orbits with self-intersections.

Note that radial motion of the T$_1$ timelike geodesics is bounded by two returning points. Then, it is in general expected that there are infinite number of self-intersections because the radial coordinate is a periodic function of $\lambda$. In these cases, there is a precession of the periastron as Fig. \ref{TL-several2} shows. However, for a certain values of $|J|-M$ the orbit is closed, and also it could contain self-intersections. To study these timelike orbits, we consider Eq. \eqref{theta(r)}, which provides the angle swept from $u_\text{min}=u_-$ to $u_\text{max}=u_+$, 
\begin{align} \label{angTL}
    |\Delta \theta| = \frac{\pi}{2 \sqrt{|J|-M}}.
\end{align}

To obtain the above expression is necessary to note that the logarithmic term in  \eqref{theta(r)} does not contribute at the returning points (where $F=\infty$) and also the property $s(u_\text{min})=-s(u_\text{max})$ which doubles and not vanish the contribution of the arctan term\footnote{It is found that for $\sgn(J) E/L <1$, $s(u_\text{min})=-s(u_\text{max})=-\sgn(JL)$,  and for $\sgn(J) E/L >1$, $s(u_\text{min})=-s(u_\text{max})=\sgn(JL)$, being $\sgn(J) E/L =1$ not possible for these geodesics.}. 

The orbit will be closed if $\Delta \theta $ given in \eqref{angTL} is a rational multiple of $\pi$ (or $\pi/2$), namely if $ |\Delta \theta|= p \pi/(2 q)$, with $p/q \in \mathbb{Q}$. This occurs for $|J|-M=q^2/p^2$ and is pictured with two examples in Fig. \ref{TL_plot3}. The orbit passes through $2n$ maximum radii, without self-intersections, provided $p/q=1/n$, namely if $\sqrt{|J|-M}=n$, as the Fig. \ref{TL_plot2} describes.  

Similar orbits were found for an effective harmonic oscillator obtained in a cosmic string background \cite{Luis-Mikhail}, where those orbits correspond to a particle under a quadratic potential, which mimics a negative cosmological constant in three dimensions.

\section{Concluding remarks} \label{sec5} 

The overspinning member of the BTZ family of spacetimes has been studied in this article. Although announced 30 years ago \cite{BTZ2}, this locally AdS$_3$ spacetime had not received much attention until now. As the other members of the BTZ family, the overspinning solution can be obtained by identifications in the covering pseudosphere in $\mathbb{R}^{2,2}$, as explicitly shown in Sec.  \ref{identificaciones}. Moreover, it contains a naked singularity at the origin \cite{BMZ2} which prevents the extension of the spacetime to the region $r^2 \le 0$, which contains closed timelike curves. As shown in Sec.  \ref{noCTC}, there are no closed timelike curves if $r^2$ is restricted to positive values. This refutes the claim \cite{BGG,Compere:2019qed} which states that the overspinning spacetimes contain closed timelike curves. Thus, the entire BTZ family is free of causal paradoxes.

To achieve a better analysis of the geodesics in the overspinning solution, a review of the geodesics for the complete BTZ family was included. The geodesic equations were integrated in terms of elementary functions and the circular geodesics were studied in detail providing new results. A classification taking into account the radial bounds leads to eight types of noncircular geodesics for this family of spacetimes.  In the case of the overspinning naked singularity, null and spacelike geodesics are classified in three types: (i) those that can reach infinity passing by a point nearest to the singularity, (ii) others extend from the central singularity to infinity, and (iii)  those with a radial upper bound and terminating at the singularity.  The timelike geodesics cannot reach infinity, as expected for an anti-de Sitter spacetime. They either loop around the singularity or fall into it. The spatial projections of the geodesics have self-intersections. The number of self-intersections in null and spacelike geodesics was calculated in Sec. \ref{autointer}. It was found that the orbits of timelike geodesics are closed if $\sqrt{|J|-M}$ is a rational number, and have no self-intersections if it is a natural number.

One question that can be asked is, given a geodesic in a three dimensional space with negative cosmological constant, is it possible to discern the type of spacetime within the BTZ family? Is it possible to determine the values of $M$ and $J$ from the examination of the geodesics? In particular, given an asymptotically AdS$_3$ spacetime, can an experiment be devised to determine whether it corresponds to a black hole, a conical or an overspinning geometry? We leave these questions to be treated in future work.

It has been shown that a BTZ black hole can be formed by a gravitationally collapsing dust  \cite{Ross:1992ba} or by a null fluid \cite{Husain:1995yb} in AdS$_3$. A similar collapse that could result the overspinning naked singularity seems unlikely. In a future work \cite{BMZ2} we are going to study the problem of the source in this spacetime.

It has been recently observed that the partition function of AdS$_3$ gravity may require the inclusion of conical geometries in the spectrum \cite{Benjamin-et-al}, presumably improving the consistency of previous results that only include black hole geometries \cite{Maloney-Witten}. The overspinning spacetime nature is similar to the black hole and conical geometries and completes the spectrum of values $(M,J)\in \mathbb{R}^2$, which suggests that these geometries could also be justifiably included in the partition function for AdS$_3$ gravity.

Another recent work \cite{ISCO} studies the innermost stable circular orbits in $D\geq 3$ dimensions for static AdS$_D$ black holes. There, the existence of a relationship is argued between those orbits and certain metastable states on the dual CFT, which do not thermalize on a typical thermal time scale. The same analysis for circular orbits in the rotating three dimensional geometries discussed here could also give some clues about a dual CFT theory.

\section*{Acknowledgements}
We thank Luis Inzunza, Hideki Maeda, Nicol\'as Parra and Nicol\'as Vald\'es for useful discussions. This work has been partially funded by ANID/FONDECYT grants 1180368 and 1201208. 

\appendix
\section{Types of geodesics according to the constants of motion in the overspinning spacetime} \label{apendice}

\subsection{Null geodesics ($\varepsilon=0$)}
As shown in Table \ref{table-radial-bounds}, for $\varepsilon =0$ there are three types of null geodesics. In two of them (N$_1$ and N$_2$), the geodesics reach infinity but only those of type N$_2$ end at $r=0$. The null geodesics of type N$_3$ have a radial upper bound and terminate at $r=0$. The constants $B$ and $C$ reduce in the null case to $B=E^2-L^2$ and $C=-L^2 M +JEL$. Then, if $L=0$, we have $B>0$ and $C=0$, which corresponds to geodesics of type  N$_2$. Next, $L, E \neq 0$ is considered. Type N$_1$ requires the condition $\sgn(J) E/L>1$, while type N$_2$ needs $\sgn(J) E/L<  -1$. Note that $B$ and $C$ cannot vanish simultaneously for null geodesics in the overspinning geometry. The third type, N$_3$, arises if $-1< \sgn(J) E/L < M/|J|<1$ for $E\neq 0$. For a vanishing $E$, $B= -L^2$ and $C=-L^2 M$ must be negative for the existence of a geodesic, i.e., $M$ should be positive. In fact, $E=0$ null geodesics belong to  N$_3$ type with an upper bound given by $u_{\text{max}}=M$.

\subsection{Timelike geodesics ($\varepsilon=1$)}
There are two types of timelike geodesics in the overspinning spacetime. While both types have a radial upper bound,  only T$_1$ geodesics possess a nonvanishing lower bound. Type T$_2$ geodesics end at $r=0$. Since these types have opposite sign of $C$ as shown in Table \ref{table-radial-bounds}, it is convenient to consider $C$ as a quadratic function of $L$ with roots given by
\begin{equation} \label{Lpm}
L_{\pm}=\frac{J E\pm |J|\sqrt{E^2+\varepsilon M}}{2M},
\end{equation}
while for $L=0$, $C$ reduces to $J^2/4>0$.

Type T$_1$, is defined in Table \ref{table-radial-bounds} by the conditions $C>0, B>0$ and $\Delta >0$. The first condition is always fulfilled for $L=0$ and the remaining two conditions imply $E^2>|J|-M$ in this case. For $L\neq 0$, the condition $C>0$ depends on $M$ as follows: for $M>0$, $L$ must be in the range $(L_-,L_+)$, while for $M<0$, either $M+E^2 \geq 0$ with $L$ outside the interval $(L_+,L_-)$, or  $M+E^2<0$, in which case $C$ is always positive. Lastly, for $M=0$, $C>0$ requires $J(E L+ J/4)>0$. 
The condition $B>0$ means $M+E^2>L^2$. The third condition for this type is $16\Delta=B^2-4C >0$. Since $J^2>M^2$ it follows from  \eqref{radius-delta} that $\Delta>0$ implies  $E^2+L^2+M > |J+2E L|$.

The second type T$_2$ accommodates geodesics that reach $r=0$ which occur if $C \leq 0$. This condition does not hold if $L=0$ and for $L\neq 0$ it depends on $M$: if $M>0$, $L$ must satisfy $L_-\leq L \leq L_+$. Type T$_2$ is not possible for $M+E^2<0$ and for $M+E^2 \geq 0$ with $M<0$, $L$ must be in the interval $[L_+,L_-]$. For $M=0$ the condition $C\leq 0$ reduces to $J(E L+ J/4)\leq 0$, which requires $JEL<0$. Note that T$_2$ type does not accommodate timelike geodesics with $M=0$ and $E=0$. 

\subsection{Spacelike geodesics ($\varepsilon=-1$)}
Spacelike geodesics of type S$_1$ extend from $r=0$ to infinity. Setting $\varepsilon =-1$, they occur in the cases (i) $B=-M+E^2-L^2 \ge 0$ and $C=-L^2 M +JEL-J^2/4 \leq 0$, and (ii) $B\leq 0$ and $\Delta \leq 0$.

Let us consider case (i) with $L, E\neq 0$. Since $E^2-M \ge L^2>0$, the roots $L_\pm$ in \eqref{Lpm} are real. Therefore, the condition $C \leq 0$ holds if $L$ is outside the interval $(L_-,L_+)$ for  $M>0$, and $L_-\leq L \leq L_+$ for $M<0$. For $M=0$, this geodesics exist if $E^2>L^2$ and $J(EL-J/4)\leq 0$. For $L=0$, type S$_1$ just needs $E^2> M$ since $C=-J^2/4 <0$.  For $E=0$ we have the conditions $L^2 \leq -M$ and $-M L^2 \leq  J^2/4$, which simultaneously hold only if $M<0$. In case (ii), $\Delta \le 0$ implies $(E+\sgn(J) L)^2 \leq M+|J|$, and $E^2-L^2 \le M$. Set (ii) provides geodesics that reach a minimum velocity $\dot{u}^2_{\text{min}}=-16\Delta$ at $u=-B/2$, where $\ddot{u}=0$ and $\ddot{r}= 8 \sqrt{2}\Delta (-B)^{-3/2}$. 

Spacelike geodesics of type S$_2$ extend from $r=0$ to a finite upper bound. For these geodesics, the constant $A_0$ in \eqref{ulambda} is necessarily negative, and the conditions $B=-M+E^2-L^2 < 0$ (namely, $E^2-L^2 < M$), $C=-L^2 M +JEL-J^2/4 <0$ and $\Delta \geq 0$ must be satisfied. Let us consider $L, E\neq 0$. For $M<0$, we have $E^2-M>0$ and then $C<0$ holds if $L\in (L_+,L_-)$. In the case $M>0$ with $E^2-M \ge 0$, $L$ is outside the interval $(L_-,L_+)$, while if $E^2-M <0$, $C<0$ regardless $L$. The condition $\Delta \geq 0$ implies $(E+L)^2 \ge M+J$ and $(E-L)^2 \ge M-J$, which is possible only if $E^2+L^2 \ge -M$. For $M=0$, the previous conditions reduce to $E^2<L^2$,  $J(EL-J/4) <0$ and $(E\pm L)^2 \geq \pm J$ (both signs simultaneously). For $L=0$, type S$_2$ requires $E^2 <M$, hence $M$ must be positive and $E^2 \ge \text{min}(M\pm J)$. The condition on $C$ holds since $C=-J^2/4 <0$.  For $E=0$ we have the conditions $L^2 > -M$ and $M L^2 > -J^2/4$, which are trivially satisfied if $M>0$, but they become constraints for $M<0$. With $E=0$, the condition $\Delta \le 0$ reduces to $L^2 \ge \text{min}(M\pm J)$. 

The spacelike geodesics extending from infinity to a lower bound $u_+$ constitute type S$_3$ in Table \ref{table-radial-bounds}. There are two different sets of conditions for this type: (i)  $B<0$, $C < 0$, $\Delta \ge 0$, $A_0 >0$ and (ii) $C>0$. The set (i) is the same as for S$_2$ type described above, except for the condition $A_0 >0$ in \eqref{ulambda}. Set (ii) is just defined by $C=-L^2 M +JEL-J^2/4 >0$. As before, let us consider $L, E\neq 0$. For $M<0$, we have $E^2-M>0$ and then $C>0$ holds if $L \notin [L_+,L_-]$. In the case $M>0$ with $E^2-M \ge 0$, $L$  must range as $L \in (L_-,L_+)$, and there there is no such a geodesic if $E^2 <M$, with $M>0$.  For $M=0$, we have the condition  $J(EL-J/4) >0$. Set (ii) is not possible for $L=0$. Then, all S$_3$ geodesics with $L=0$ come from the set (i). For $E=0$ there is the requirement $M L^2 < -J^2/4$, which can be only satisfied if $M<0$.


\begin{thebibliography}{99}


\bibitem{AyonBeato:2004if}
E.~Ay\'on-Beato, C.~Mart\'{\i}nez and J.~Zanelli, \textit{Birkhoff's theorem for three-dimensional AdS gravity}, Phys. Rev. D \textbf{70}, 044027 (2004) doi:10.1103/PhysRevD.70.044027. [arXiv:hep-th/0403227 [hep-th]].

\bibitem{BTZ1} M.~Ba\~nados, C.~Teitelboim and J.~Zanelli, \textit{The black hole in three-dimensional space-time}, Phys.\ Rev.\ Lett.\  {\bf 69}, 1849 (1992).[hep-th/9204099]. doi:10.1103/PhysRevLett.69.1849

\bibitem{BTZ2} M.~Ba\~nados, M.~Henneaux, C.~Teitelboim and J.~Zanelli, \textit{Geometry of the (2+1) black hole}, Phys.\ Rev.\ D {\bf 48}, 1506 (1993). [gr-qc/9302012]. Erratum: [Phys.\ Rev.\ D {\bf 88}, 069902 (2013)
doi:10.1103/PhysRevD.48.1506, 10.1103/PhysRevD.88.069902

\bibitem{C-H} O. Coussaert and M. Henneaux, ``Selfdual solutions of (2+1) Einstein gravity with a negative cosmological constant,'' in \textit{The Black Hole, 25 Years Later}, edited by C. Teitelboim and J. Zanelli, World Scientific, Singapore, 1998, [arXiv:hep-th/9407181 [hep-th]].

\bibitem{MZ} O.~Mi\v{s}kovi\'c and J.~Zanelli, \textit{Negative spectrum of the 2+1 black hole}, Phys.\ Rev.\ D {\bf 79}, 105011 (2009). [arXiv:0904.0475 [hep-th]]

\bibitem{BMZ2} M. Brice\~no, C. Mart\'{i}nez and J. Zanelli \emph{On the central singularity of the BTZ geometries} (to appear).

\bibitem{Cruz1994} N.~Cruz, C.~Mart\'{\i}nez and L.~Pe\~na, \textit{Geodesic structure of the (2+1) black hole}, Class.\ Quant.\ Grav.\  {\bf 11}, 2731 (1994). [gr-qc/9401025].
doi:10.1088/0264-9381/11/11/014

\bibitem{Martinez:2019nor} C.~Mart\'{\i}nez, N.~Parra, N.~Vald\'es and J.~Zanelli, \textit{Geodesic Structure of Naked Singularities in AdS$_3$ Spacetime}, Phys.\ Rev.\ D {\bf 100}, no. 2, 024026 (2019). doi:10.1103/PhysRevD.100.024026. [arXiv:1902.00145 [hep-th]].

\bibitem{CFMZ} M.~Casals, A.~Fabbri, C.~Mart\'{\i}nez and J.~Zanelli, {\it Quantum-corrected rotating black holes and naked singularities in ( 2+1 ) dimensions}, Phys.\ Rev.\ D {\bf 99}, no. 10, 104023 (2019). doi:10.1103/PhysRevD.99.104023 [arXiv:1902.01583 [hep-th]].

\bibitem{Steif} A.~Steif, \emph{Supergeometry of three-dimensional black holes, Phys.Rev. D} \textbf{53} (1996) 5521-5526, doi: 10.1103/PhysRevD.53.5521 [arXiv:9504012 [hep-th]].

\bibitem{Hawking-Ellis} S. W. Hawking and G. F. Ellis, \emph{The Large Scale Structure of Spacetime} Cambridge University Press, Cambridge, U.K. 1973.  
  
\bibitem{Ellis-Schmidt} G.~F.~R. Ellis and  B.~G. Schmidt, \textit{Singular space-times}, Gen. Rel. Grav. {\bf 8}, 915 (1977) doi:10.1007/BF00759240; \textit{Classification of singular space-times}, Gen. Rel. Grav. {\bf 10}, 989 (1979) doi 10.1007/bf00776518.

\bibitem{Deser:1991ye} S.~Deser, R.~Jackiw and G.~'t Hooft, {\it Physical cosmic strings do not generate closed timelike curves}, Phys.\ Rev.\ Lett.\ {\bf 68}, 267 (1992). doi:10.1103/PhysRevLett.68.267 

\bibitem{Farina:1993xw}
C.~Farina, J.~Gamboa and A.~J.~Segui-Santonja,
\textit{Motion and trajectories of particles around three-dimensional black holes}, Class. Quant. Grav. \textbf{10}, L193-L200 (1993). doi:10.1088/0264-9381/10/11/001. [arXiv:gr-qc/9303005 [gr-qc]].

\bibitem{GR tables} I. S. Gradshteyn and I. M. Ryzhik I, \textit{Table of Integrals, Series And Products}, 7th edition, Elsevier Academic Press, 2007.

\bibitem{Luis-Mikhail} L. Inzunza and M. S. Plyushchay, \textit{Conformal bridge in a cosmic string background}, arXiv:2012.04613 [hep-th].
 
 \bibitem{BGG} G.~Barnich, A.~Gomberoff and H.~A.~Gonz\'alez, \emph{The flat limit of three dimensional asymptotically anti-de Sitter spacetimes}, Phys.\ Rev.\ D \textbf{86}, 024020 (2012). doi:10.1103/PhysRevD.86.024020 [arXiv:1204.3288 [gr-qc]]. 
 
\bibitem{Compere:2019qed} G.~Comp\`ere, \emph{Advanced Lectures on General Relativity}, Lect.\ Notes Phys.\  {\bf 952}, 150 (2019). doi:10.1007/978-3-030-04260-8; G.~Comp\`ere and A.~Fiorucci, \emph{Advanced Lectures on General Relativity}, arXiv:1801.07064 [hep-th].

\bibitem{Ross:1992ba} R.~B.~Mann and S.~F.~Ross, \textit{Gravitationally collapsing dust in (2+1)-dimensions}, Phys. Rev. D \textbf{47} (1993), 3319-3322. doi:10.1103/PhysRevD.47.3319. [arXiv:hep-th/9208036 [hep-th]].

\bibitem{Husain:1995yb} V.~Husain, \textit{Black hole solutions in (2+1)-dimensions}, Phys. Rev. D \textbf{52} (1995), 6860-6862. doi:10.1103/PhysRevD.52.6860 [arXiv:gr-qc/9511003 [gr-qc]].

\bibitem{Benjamin-et-al} N.~Benjamin, S.~Collier and A.~Maloney, \textit{Pure Gravity and Conical Defects}, JHEP {\bf 2009} (2020) 034. doi:10.1007/JHEP09(2020)034. [arXiv:2004.14428 [hep-th]].

\bibitem{Maloney-Witten} A.~Maloney and E.~Witten, \textit{Quantum Gravity Partition Functions in Three Dimensions}, JHEP {\bf 1002} (2010) 029. doi:10.1007/JHEP02(2010)029. [arXiv:0712.0155 [hep-th]].

\bibitem{ISCO} D.~Berenstein, Z.~Li and J.~Simon, \emph{ISCOs in AdS/CFT, Class. Quant. Grav.} \textbf{38} (2021) 4, 045009. doi: 10.1088/1361-6382/abcaeb [arXiv:2009.04500 [hep-th]].

\end{thebibliography}
\end{document}